\documentclass[
  reprint,
  aps,
  prb,
  amsmath,
  amssymb,
  superscriptaddress,
  longbibliography
]{revtex4-2}

\usepackage{newtxtext} 
\usepackage{newtxmath} 
\usepackage{needspace}
\usepackage{amsmath} 
\usepackage{upgreek} 
\usepackage{bbm} 
\usepackage{graphicx} 
\usepackage{dcolumn} 
\usepackage{bm} 
\usepackage{braket}
\usepackage[percent]{overpic}

\usepackage[colorlinks]{hyperref} 
\hypersetup{
    colorlinks=true,    
    linkcolor=blue,     
    citecolor=blue,     
    urlcolor=blue       
}

\makeatletter
\def\@bibdataout@revtex#1#2#3#4{%
  \immediate\write\@bibdataout{%
    @article{#1,
      author = {#2},
      title = {#3},
      journal = {#4},
      url = {\@bibdata@url},
      doi = {\@bibdata@doi}
    }%
  }%
}
\makeatother

\newcommand{\img}{\mathrm{i}} 

\begin{document}

\title{Composite-Fermion Study of Cavity-Modified Fractional Quantum Hall Excitation Gaps}

\author{Dalin Bori\c{c}i}
\affiliation{Universit\'{e} Paris Cit\'e, CNRS, Mat\'{e}riaux et Ph\'{e}nom\`{e}nes Quantiques, 75013 Paris, France}

\author{Nicolas Regnault}
\affiliation{Center for Computational Quantum Physics, Flatiron Institute, 162 5th Avenue, New York, NY 10010, USA}
\affiliation{Laboratoire de Physique de l’Ecole Normale Sup\'erieure, ENS, Universit\'e PSL, CNRS, Sorbonne Universit\'e, Paris, France}
\affiliation{Department of Physics, Princeton University, Princeton, NJ 08544, USA}

\author{ Cristiano Ciuti}
\affiliation{Universit\'{e} Paris Cit\'e, CNRS, Mat\'{e}riaux et Ph\'{e}nom\`{e}nes Quantiques, 75013 Paris, France}

\date{\today}

\begin{abstract}
\noindent
We investigate how cavity-mediated attractive electron-electron interactions modify the excitation gaps of fractional quantum Hall states within the composite-fermion framework. We compute both the neutral magnetoroton excitation spectrum and the charged excitation gap relevant to transport experiments for the Laughlin $\nu=1/3$ and $\nu=1/5$ states. We consider a spin-polarized lowest-Landau-level model in which the interaction is mediated by a cavity mode with a spatially uniform vacuum-field gradient and a finite interaction range controlled by a long-distance cutoff. Finite-size scaling reveals that the transport gap is consistently enhanced by the cavity-induced interaction, with the gap enhancement scaling quadratically with the electron number and with the fourth power of the vacuum-field gradient. By contrast, the magnetoroton spectrum exhibits a richer dependence on the interaction range. The high-$k$ magnetoroton gap is enhanced for all interaction ranges considered, consistent with its close connection to the charged excitation gap, even with the long-range character of the interaction.

\end{abstract}

\maketitle

\section{\label{sec:intro}
    Introduction
}

Fractional quantum Hall liquids
are paradigmatic examples of strongly correlated electronic systems~\cite{TsuiStormer}. 
In a partially filled Landau level,
the kinetic energy is quenched by the flat-band dispersion,
and the electronic properties are governed entirely
by electron–electron interactions.
The resulting incompressible quantum liquids
exhibit finite excitation gaps of purely many-body origin~\cite{Laughlin_wf, HaldanePseudoPotentials, jainBook}.
For the Coulomb interaction and the simplest fractional quantum Hall states,
finite-size calculations show that these gaps depend on the number of electrons
but converge rapidly toward their thermodynamic-limit values; systems containing only about ten electrons already provide a good approximation to the large-system limit~\cite{HaldaneRezayi_ED, FanoOrtolaniColombo, Morf_ED}.

More recently, cavity quantum electrodynamics has emerged
as a promising platform for engineering
and controlling many-body electronic systems
through their coupling to the quantized electromagnetic field.
In this rapidly developing field of cavity-altered quantum materials,
vacuum fluctuations and virtual photon exchanges can modify effective interactions,
collective excitations, and phase stability
even in the absence of external driving~\cite{
    GarciaVidal2021,
    Schlawin2022,
    Bloch2022,
    Lorenzo_review,
    lu2025cavity}.
The prospect of tailoring material properties through cavity-mediated interactions has stimulated intense theoretical and experimental activity across a broad range of systems,
including superconductors,
magnetic materials,
excitonic and polaritonic platforms,
and topological quantum matter~\cite{
    Enkner,
    appugliese_breakdown,
    LorenzoStripes,
    SambuddhaStripes,
    imamoglu2026,
    basov2025polaritonic,
    Ciuti2021,
    Nguyen2023,
    perez2023light,
    Scalari2012,
    Keller2017,
    ParaviciniBagliani2018,
    Ashida2023,
    Kuroyama2023,
    Kuroyama2024,
    HonkKong,
    Jarc2023,
    Sentef2018,
    Hagenmller2017,
    Hagenmller2018,
    Arwas2023,
    Winter2025,
    Macedo2024,
    Borici2025,
    Li2022,
    Becerra2025,
    Buonemani2026,
    RitzZwilling2026,
    Lin2023,
    mendez2023edge,
    Dmytruk2022,
    nguyenSSH,
    nguyen2026strange,
    dmytruk2024hybrid,
    gomez2024high,
    shaffer2024entanglement,
    bacciconi2024topological,
    yang2024emergent,
    miguel,
    bacciconi2024theory,
    dag2024engineering,
    Jaksch_sc,
    Basov_sc,
    Curtis_sc,
    Klinovaja_sc}.

Fractional quantum Hall systems constitute a particularly attractive setting for exploring such effects.
A recent work~\cite{
    Enkner}
has shown experimentally cavity-modified fractional quantum Hall gaps
via quantum magneto-transport measurements on a two-dimensional GaAs electron system
embedded in a split-ring electromagnetic resonator.
In the same paper, it was shown that a cavity mode
with a spatially varying vacuum field can mediate an effective
intraband electron–electron interaction through the exchange of virtual cavity photons,
where intraband refers to processes occurring within the same Landau level.
The resulting interaction is attractive
and acts between electrons located in regions where the vacuum field
exhibits a finite spatial gradient.
Due to the complexity of the problem,
the theoretical approach considered a simplified scenario
of interaction mediated by a cavity mode with a uniform spatial gradient.
This framework was used to investigate cavity-induced modifications
of spin splitting at odd integer filling factors,
which can be interpreted as a renormalization of the electronic g-factor,
as well as cavity-induced changes of the excitation gap of the Laughlin~\cite{
    Laughlin_wf}
fractional quantum Hall state.
In particular, the neutral magnetoroton spectrum
was studied using the Girvin–MacDonald–Platzman (GMP) single-mode approximation~\cite{
    GMP}
with a cutoff regularization.

In the present work, we address this problem within the composite-fermion (CF) framework~\cite{
    JainCF89},
which provides a quantitatively more accurate description of fractional quantum Hall excitations~\cite{
    Scarola_CFE,
    Bonesteel_quasiElectron}.
In addition to the neutral magnetoroton spectrum,
we determine the charged excitation gap,
the quantity most directly relevant to transport experiments~\cite{bonesteel1995composite}.
Using a systematic finite-size scaling analysis,
we investigate how cavity-mediated interactions modify this gap as a function of particle number
and of the characteristic length scales of the effective interaction.
We find that the resulting enhancement scales as the fourth power of the vacuum-field gradient
and quadratically with the electron number,
revealing a strongly collective effect that contrasts sharply with the finite-size behavior
of conventional Coulomb-interacting fractional quantum Hall systems.

This article is organized as follows.
In Sec.~\ref{sec:theory},
we introduce the theoretical framework, including the effective Hamiltonian with the cavity-mediated electron-electron interaction, the composite-fermion formalism, and the definitions of the neutral and charged excitations together with their corresponding excitation gaps.
In Sec.~\ref{sec:Results},
we present our numerical results for the cavity-induced modifications of the charged excitation gap and the neutral excitation spectrum, together with a finite-size scaling analysis of the relevant observables.
Finally, conclusions and perspectives are presented in Sec.~\ref{sec:conclusions}. 
Technical details are deferred to the Appendices.

\section{\label{sec:theory}
    Theoretical Framework
}

In this section, we present the theoretical framework used throughout this work.
We begin by introducing the microscopic cavity quantum electrodynamics model and the corresponding effective cavity-mediated electron-electron interaction within the lowest Landau level.
Next, we present the composite-fermion wavefunctions used to describe the ground, neutral excited, and charged excited states.
Finally, we describe the numerical approach employed to evaluate the corresponding many-body excitation energies.

\subsection{Quantum light-matter model and effective interaction}

We consider the same cavity quantum electrodynamics model introduced in Ref.~\cite{Enkner} for a spin-polarized fractional quantum Hall system coupled to a single quantized cavity mode.
The electrons are confined to a two-dimensional plane and subjected to a uniform perpendicular magnetic field.
When the lowest Landau level (LLL) is partially filled, Landau quantization quenches the kinetic energy, so that the low-energy physics is governed entirely by electron-electron interactions projected onto the LLL.

The microscopic starting point is the minimal coupling Hamiltonian
\begin{align}\label{eqn:startingHamiltonian}
    \nonumber
    \hat{H}
    &=
    \frac{1}{2m}
    \sum_i
    \left[
        \hat{\pi}_{\mu}^{(i)}
        +
        e\hat A_{\mu}^{(\rm cav)}
        \!\left(
        \hat r^{(i)}
        \right)
    \right]
    \left[
        \hat{\pi}_{\mu}^{(i)}
        +
        e\hat A_{\mu}^{(\rm cav)}
        \!\left(
        \hat r^{(i)}
        \right)
    \right]
    \\
    &\qquad\qquad+
    \hat V_{\rm C}
    +
    \hbar\omega_{\rm cav}
    \hat{\mathcal A}^{\dagger}
    \hat{\mathcal A} \, ,
\end{align}
where $-e$ is the electron charge 
and $m$ the electronic band mass.
Here,
\begin{equation}
    \label{eqn:kineticMomentum}
    \hat{\pi}_{\mu}
    =
    \hat p_{\mu}
    +
    eA_{\mu}
\end{equation}
is the kinetic momentum associated with the static magnetic field
\footnote{Greek indices label the two spatial directions $x$ and $y$,
and repeated indices are summed over},
$\hat V_{\rm C}$ is the Coulomb interaction,
$\hat{\mathcal A}^{\dagger}$ creates a cavity photon with frequency
$\omega_{\rm cav}$,
and $\hat A_{\mu}^{(\rm cav)}$
is the in-plane component of the cavity vector potential,
given by
\begin{equation}\label{eqn:cavityVectorPotential}
    \hat A_{\mu}^{(\rm cav)}
    \!\left(
    \hat r
    \right)
    =
    A_0
    \left[
    \delta_{\mu x}
    +
    \mathcal G
    \hat r_x
    \delta_{\mu y}
    \right]
    \left(
    \hat{\mathcal A}
    +
    \hat{\mathcal A}^{\dagger}
    \right) \, ,
\end{equation}
where $A_0$ sets the overall vacuum-field amplitude
and $\mathcal G$ is the spatial gradient of the cavity mode.

A spatially uniform cavity field couples only to the cyclotron motion and therefore cannot modify the guiding-center dynamics within a single Landau level, in accordance with Kohn's theorem~\cite{kohnsTheorem, rokajWeakened}.
A finite spatial gradient is therefore the {\it minimal} ingredient required to generate cavity-mediated electron-electron interactions within the same Landau level. Following Ref.~\cite{Enkner}, we adopt the simplest realization of this mechanism by assuming a spatially uniform cavity-field gradient, as described by Eq.~\eqref{eqn:cavityVectorPotential}.
Although the electromagnetic field profile of a realistic cavity is generally more complex, this minimal model already captures the essential mechanism responsible for cavity-mediated interactions while remaining amenable to quantitative many-body calculations.

Projecting the Hamiltonian onto the LLL and subsequently eliminating the photonic degrees of freedom through a Schrieffer-Wolff transformation~\cite{SW}, we obtain an effective electron-electron interaction whose pair potential is
\begin{equation}
\label{eqn:harmonicsCavPlaneCut}
    V_{\rm cav}(r;\mathcal L)
    =
    -\xi
    \left(
    \frac{(r/\ell)^4}{16}
    -
    (r/\ell)^2
    +
    2
    \right)
    e^{-r^2/\mathcal L^2},
\end{equation}
where
\begin{align}
\label{eqn:xi}
    \xi
    &\equiv
    \frac{
        3 e^4 (\mathcal{G}_{\rm E} \ell)^4
    }{
        8 m^2 \hbar \omega_{\rm cav}^5
    }
    \, ,
\end{align}
$\ell \equiv \sqrt{\hbar/eB}$ is the magnetic length
and $\mathcal{G}_{\rm E} \equiv \omega_{\rm cav} A_0 \mathcal{G}$ is the gradient of the cavity electric field
\footnote{
    We note that the gradient parameter $\mathcal{G}$
    has dimensions $\,{\rm L}^{-1}$ and is related to the
    vector potential gradient parameter $\mathcal{G}_{\rm A}$ from Ref.~\cite{Enkner}
    and the experimentally relevant $\mathcal{G}_{\rm E}$ as:
    \begin{align*}
        \mathcal{G}_{\rm A}
        =
        A_0 \mathcal{G} \, ,
        \qquad
        \mathcal{G}_{\rm E}
        =
        E_{\rm vac} \mathcal{G} \, ,
    \end{align*}
    with $E_{\rm vac} = A_0 \omega_{\rm cav}$.
}.
The Gaussian envelope introduces a characteristic interaction range $\mathcal L$, which phenomenologically accounts for the finite spatial extent over which a realistic cavity mode exhibits appreciable field gradients.
The limiting case $\mathcal L\rightarrow\infty$ recovers the idealized model of Ref.~\cite{Enkner}, corresponding to a spatially uniform cavity-field gradient throughout the system.
A detailed derivation of the effective interaction is presented in App.~\ref{app:theory}.
Throughout this work, we denote the corresponding LLL-projected interaction by $\hat V_{\rm cav}^{\mathcal L}$.

It is useful to interpret the structure of Eq.~\eqref{eqn:harmonicsCavPlaneCut}
in terms of Haldane pseudopotentials~\cite{HaldanePseudoPotentials}.
In the untruncated limit, $\mathcal{ L}\to\infty$,
the cavity-mediated interaction gives pseudopotential components of the form
\begin{equation}
    v_m^{\rm (cav)}
    =
    -\xi (m^2-m) .
\end{equation}
Therefore, for the leading fermionic short-range channel $m=1$,
the cavity contribution vanishes, $v_1^{\rm (cav)}=0$.
As a result, the cavity-mediated interaction leaves the Coulomb
pseudopotential $v_1^{\rm (C)}$ unchanged.
Note that this is the dominant component responsible for opening
the Laughlin gap at filling factor $\nu=1/3$.
By contrast, for $m>1$, the attractive cavity-mediated contribution
reduces the longer-range Coulomb pseudopotentials
$v_{m>1}^{\rm (C)}$.
This is particularly significant because the Laughlin gap of the
idealized $V_1$ model,
defined by the pseudopotentials
$V_m=V_1^{\rm (C)}\delta_{m,1}$,
is known to be larger than that of the full Coulomb interaction.
Since the longer-range Coulomb pseudopotentials
$v_{m>1}^{\rm (C)}$ reduce the gap generated by
$v_1^{\rm (C)}$, the structure of the cavity pseudopotentials
already suggests that adding the attractive cavity-mediated interaction
on top of the repulsive Coulomb interaction can enhance the Laughlin gap,
thereby strengthening the incompressibility of the Laughlin state.
We will show in Sec.~\ref{sec:Results},
using composite fermion theory,
and in App.~\ref{app:ED_benchmarks}, using exact diagonalization,
that this is indeed the case.

\subsection{Composite-fermion ansatz}\label{sec:composite}

To compute the fractional quantum Hall excitation gaps and their cavity-induced modifications, we employ composite-fermion trial wavefunctions for both the ground and excited states.
These wavefunctions provide an accurate description of the low-energy spectrum of Coulomb-interacting electrons in the lowest Landau level~\cite{jainBook}.

The composite-fermion construction maps the strongly interacting electron problem in a high magnetic field onto a weakly interacting problem of composite fermions moving in a reduced effective magnetic field~\cite{JainCF89}.
Each electron binds to $2p$ flux quanta, partially screening the external magnetic field, so that the effective magnetic field experienced by the composite fermions is
$
    B^\ast = B - 2p n_{\rm e} \phi_0 .
$
Here, $n_{\rm e}$ is the electron density and $\phi_0 = h/e$ is the magnetic flux quantum.
Equivalently, using $\nu = n_{\rm e}\phi_0/B$ and $\nu^\ast = n_{\rm e}\phi_0/B^\ast$, the electronic and composite-fermion filling factors satisfy
$
    \nu = \nu^\ast / (2p\nu^\ast+1) .
$

The Laughlin ground state at $\nu=1/(2p+1)$, denoted by $\psi_{\rm GS}$, corresponds to a completely filled lowest effective Landau level, or $\Lambda$ level.
It is therefore an integer quantum Hall state of composite fermions with $\nu^\ast=1$.

In this work, we consider both neutral and charged excitations above this ground state.
The neutral excitations are composite-fermion excitons obtained by promoting a composite fermion from the filled lowest $\Lambda$ level to the lowest empty one, producing a state with a well-defined wave vector $k$,
$
    \psi_k^{\rm exc} \, .
$
The charged excitations correspond to the composite-fermion quasihole and quasielectron states.
A quasihole is created by removing a composite fermion from the filled lowest $\Lambda$ level, whereas a quasielectron is obtained by placing one composite fermion in the lowest empty $\Lambda$ level.
We denote these states by
$
    \psi_{\rm qh}
$
and
$
    \psi_{\rm qe},
$
respectively.
Details of the composite-fermion wavefunctions and the spherical geometry used throughout this work are provided in App.~\ref{app:composite}.

\subsection{Computation of many-body energies}\label{sec:compuingEnergies}

Having introduced the effective interaction and the many-body trial wavefunctions, the remaining task is to evaluate their interaction energies.
For a given composite-fermion wavefunction $\psi$, the expectation value of the interaction Hamiltonian is
\begin{equation}
    \bra{\psi}
        \hat{V}
    \ket{\psi}
    =
    \frac{
        \int {\rm d}^{2N}r \,
            \sum_{i<j}V\left( r^{(i)} - r^{(j)} \right)
        \,
        \left|
            \psi(\{r^{(i)}\}_i)
        \right|^2
    }{
        \int {\rm d}^{2N}r \,
        \left|
            \psi(\{r^{(i)}\}_i)
        \right|^2
    }
    \, ,
\end{equation}
where $V(r)=V_{\rm C}(r)+V_{\rm cav}(r;\mathcal{L})$ is the total pair interaction potential.

The standard approach is to evaluate these expectation values using
Metropolis-Hastings Monte Carlo integration~\cite{Metropolis,Hastings},
sampling configurations with probability proportional to
$
    \left|
        \psi(\{r^{(i)}\}_i)
    \right|^2 .
$
However, this approach becomes inefficient when the interaction depends on the continuously tunable range parameter $\mathcal{L}$.
A direct implementation would require an independent Monte Carlo calculation of
$
    \bra{\psi}
        \hat{V}_{\rm cav}^{\mathcal{L}}
    \ket{\psi}
$
for every value of $\mathcal{L}$, resulting in a substantial computational overhead.
Moreover, Haldane pseudopotentials are not readily incorporated into real-space Monte Carlo calculations because of their singular real-space representation.

To overcome these difficulties, we employ the method introduced in Ref.~\cite{Mross}, which expresses the interaction in terms of its angular-momentum harmonics.
This approach has previously been used to investigate phase diagrams as a function of the interaction potential.
For each many-body state, we first compute the angular-momentum-resolved pair density, which is independent of the interaction parameters.
The interaction energy for any choice of $\mathcal{L}$ is then obtained by contracting this pair density with the corresponding harmonic components of the pair potential.
Further technical details are provided in App.~\ref{app:compuingEnergies}.

\section{Numerical Results}
\label{sec:Results}

In this section, we present our numerical results for the cavity-induced modifications of the excitation spectrum of the Laughlin fractional quantum Hall states.
We first investigate the charged excitation gap, which is directly relevant to transport experiments, and establish its finite-size scaling behavior.
We then turn to the neutral excitation spectrum, highlighting the similarities and differences in the cavity-induced modifications of charged and neutral excitations.

\subsection{Cavity-induced modification of the charge gap}

We begin by analyzing the cavity-induced correction to the charged excitation gap of the Laughlin fractional quantum Hall states.
Within the composite-fermion framework, this correction is given by
\begin{align}\label{eqn:chGap_shift}
\nonumber
\delta\Delta^{(\nu)}_{\rm ch}
&=
\bra{\psi^{\rm qe}}
\hat{V}_{\rm cav}^{\mathcal{L}}
\ket{\psi^{\rm qe}}
+
\bra{\psi^{\rm qh}}
\hat{V}_{\rm cav}^{\mathcal{L}}
\ket{\psi^{\rm qh}}
\\
&\quad\quad\quad\quad\quad\quad\quad\quad\quad\quad
-
2
\bra{\psi_{\rm GS}}
\hat{V}_{\rm cav}^{\mathcal{L}}
\ket{\psi_{\rm GS}}
\, ,
\end{align}
where the three matrix elements represent the cavity-induced energy shifts of the quasielectron, quasihole, and ground states, respectively.
\footnote{
Throughout this work, energies corresponding to different flux sectors are compared at fixed spherical radius. In our simulations, we choose
$\psi^{\rm qe} = \psi^{\rm qe}_{m_e=N/2}$
and
$\psi^{\rm qh} = \psi^{\rm qh}_{m_h=N/2}$,
since the energies are independent of the orbital angular momentum quantum number.
}
\begin{figure}[t]
    \centering
    \includegraphics[width=\columnwidth]{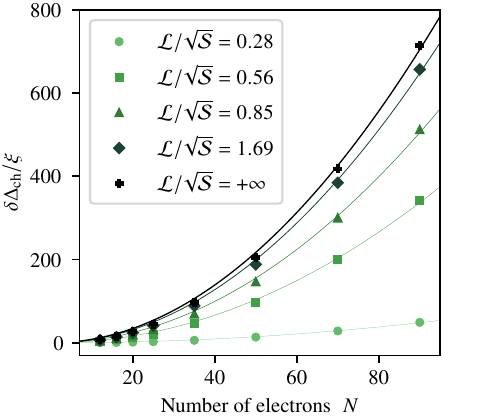}
  
\caption{
Cavity-induced correction to the charge gap,
$\delta\Delta_{\mathrm{ch}}/\xi$,
normalized to the characteristic interaction energy scale $\xi$ defined in Eq.~(\ref{eqn:xi}),
which is proportional to the fourth power of the cavity-mode spatial gradient.
The correction is shown as a function of the number of electrons $N$
for the $\nu=1/3$ Laughlin state and for several values of the dimensionless interaction range
$\mathcal{L}/\sqrt{\mathcal{S}}$.
Symbols denote composite-fermion calculations, while solid lines are global quadratic fits of the form
$
\delta\Delta_{\mathrm{ch}}
=
A_{1/3}(\mathcal{L}/\sqrt{\mathcal{S}})\,\xi N^2,
$
where the coefficient
$A_{1/3}(\mathcal{L}/\sqrt{\mathcal{S}})$
depends on the interaction range.
The increase of $A_{1/3}$ with $\mathcal{L}/\sqrt{\mathcal{S}}$
demonstrates that the cavity-induced enhancement of the charge gap becomes stronger as the interaction range increases.
}

    \label{fig:chargeGapShiftsInfCutOff_vs_N_manyLs}
\end{figure}

In Fig.~\ref{fig:chargeGapShiftsInfCutOff_vs_N_manyLs},
we show the charge-gap shift, reported in units of $\xi$,
as a function of the number of electrons $N$ at filling factor $\nu=1/3$.
Here, $\xi$ is the energy scale defined in Eq.~(\ref{eqn:xi}),
which scales as the fourth power of the cavity vacuum-field gradient.
The data are shown for several values of the interaction range $\mathcal{L}$,
expressed in units of the linear system size $\sqrt{\mathcal{S}}$,
where $\mathcal{S}$ denotes the surface area of the system.
Note that $\mathcal{S}$ varies with $N$ to keep $\nu$ fixed.
The value of $\mathcal{L}$ is scaled accordingly.
As discussed in Sec.~\ref{sec:theory},
the parameter $\mathcal{L}$ phenomenologically represents the spatial extent
over which the cavity field exhibits appreciable gradients.
It is therefore natural to assume that this characteristic length
scales proportionally with the linear size of the sample.
Physically, this corresponds to enlarging the cavity together with the electronic system,
so that the region where the cavity field exhibits significant gradients
occupies the same fraction of the sample.
 
The resulting correction to the charge gap is positive,
demonstrating that the cavity-mediated interaction
enhances the incompressibility of the fractional quantum Hall state.
Unlike the Coulomb contribution,
which rapidly approaches its thermodynamic-limit value with increasing $N$,
the cavity-induced correction grows quadratically with the number of electrons, $\delta\Delta_{\rm ch}\propto N^2$.
The quadratic scaling is confirmed by the very good agreement between the data
and the quadratic fits shown in Fig.~\ref{fig:chargeGapShiftsInfCutOff_vs_N_manyLs}
for all interaction ranges considered.
This scaling reflects the collective character of the cavity-mediated interaction,
which couples electrons over a finite fraction of the system.
Consequently, a conventional thermodynamic extrapolation
at fixed interaction strength $\xi$ is not appropriate.
We therefore adopt a finite-size scaling prescription
in which the scaled coupling $\xi N^2$ is kept fixed as $N$ is varied.
We implement this scaling
by comparing systems at fixed filling factor $\nu$
and fixed dimensionless ratio $\mathcal{L}/\sqrt{\mathcal{S}}$.

The data in Fig.~\ref{fig:chargeGapShiftsInfCutOff_vs_N_manyLs} are accurately described by the scaling law
\begin{equation}\label{eqn:scalingLaw}
    \delta\Delta^{(\nu)}_{\rm ch}
    =
    A_{\nu}\left(\mathcal{L}/\sqrt{\mathcal{S}}\right)\,
    \xi N^2 \, .
\end{equation}
The dimensionless coefficient $A_{\nu}$ depends on both the fractional quantum Hall state and the scaled interaction range $\mathcal{L}/\sqrt{\mathcal{S}}$.
Its value is obtained from quadratic fits to the numerical data shown in Fig.~\ref{fig:chargeGapShiftsInfCutOff_vs_N_manyLs}.

\begin{figure}[t!]
    \centering
    \includegraphics[width=\columnwidth]{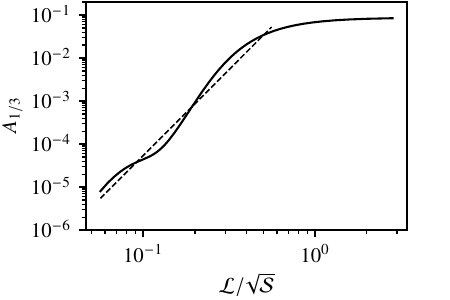}

\caption{
Dependence of the dimensionless coefficient $A_{1/3}$, which quantifies the cavity-induced enhancement of the charge gap according to Eq.~(\ref{eqn:scalingLaw}), on the scaled interaction range $\mathcal{L}/\sqrt{\mathcal{S}}$, shown on a log-log scale.
The values of $A_{1/3}$ are obtained from the quadratic fits to the finite-size scaling data reported in Fig.~\ref{fig:chargeGapShiftsInfCutOff_vs_N_manyLs}.
The dashed line is a quartic fit, highlighting the strong increase of the charge-gap enhancement with the interaction range.
}
    \label{fig:A_13_of_L}
\end{figure}

For the $\nu=1/3$ state, the coefficient $A_{\nu}$ increases monotonically with the scaled interaction range $\mathcal{L}/\sqrt{\mathcal{S}}$, as shown in Fig.~\ref{fig:A_13_of_L}.
This reflects the increasingly extended character of the cavity-mediated interaction.
In the limit $\mathcal{L}\gg\sqrt{\mathcal{S}}$, the interaction effectively spans the entire system, producing the largest enhancement of the charge gap.

\begin{figure}[t!]
    \centering
    \includegraphics[width=\columnwidth]{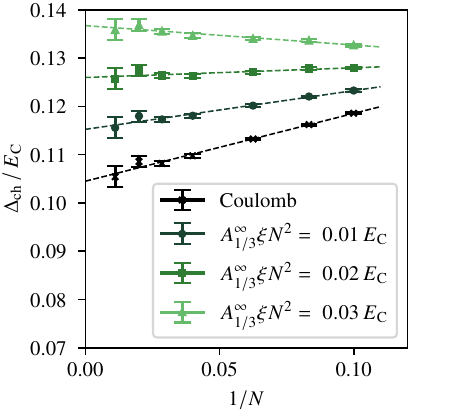}

\caption{
Thermodynamic extrapolation of the charge gap of the Laughlin $\nu=1/3$ state for the limiting case
$\mathcal{L}/\sqrt{\mathcal{S}}\rightarrow+\infty$.
The charge gap is plotted as a function of $1/N$ and linearly extrapolated to the thermodynamic limit.
To obtain a meaningful extrapolation of the cavity-modified gap, the interaction strength is scaled so that $\xi N^2$ remains constant as $N$ is varied.
The cavity coupling is parameterized by the scaled energy
$A_{1/3}^{\infty}\xi N^2$, where
$
A_{1/3}^{\infty}
\equiv
A_{1/3}(\mathcal{L}/\sqrt{\mathcal{S}}\rightarrow+\infty)
=
0.087
$
is obtained from the finite-size scaling analysis shown in Fig.~\ref{fig:chargeGapShiftsInfCutOff_vs_N_manyLs}.
The extrapolated results demonstrate that the cavity-induced enhancement of the charge gap remains finite in the thermodynamic limit.
}
    \label{fig:cavity_tl_CF}
\end{figure}

Having established the finite-size scaling law,
we now extrapolate the cavity-modified charge gap
to the thermodynamic limit while keeping both $\xi N^2$ and $\mathcal{L}/\sqrt{\mathcal{S}}$ fixed.
Figure~\ref{fig:cavity_tl_CF} shows this extrapolation for the limiting case
$\mathcal{L}/\sqrt{\mathcal{S}}\rightarrow+\infty$.
The charge gap is plotted as a function of $1/N$
for the bare Coulomb interaction and for several fixed values
of the scaled coupling $A_{1/3}^{\infty}\xi N^2$,
with $A_{1/3}^{\infty} \equiv A_{1/3}(\mathcal{L}/\sqrt{\mathcal{S}} \to \infty)$.
For the bare Coulomb problem, our results are in excellent agreement with those of Ref.~\cite{Bonesteel_quasiElectron}.
The extrapolated gaps demonstrate that the cavity-induced enhancement persists in the thermodynamic limit, with its magnitude determined by the scaled energy
$A_{1/3}^{\infty}\,\xi N^2$.
In App.~\ref{app:ED_benchmarks}, we repeat this analysis for the smaller system sizes accessible to exact diagonalization (ED), obtaining consistent thermodynamic extrapolations and excellent agreement with the composite-fermion calculations.

We have performed the same finite-size scaling analysis for the $\nu=1/5$ Laughlin state and find the same scaling behavior as for $\nu=1/3$.
In particular, the cavity-induced correction to the charge gap scales quadratically with the number of electrons,
$\delta\Delta_{\rm ch}\propto N^2$,
with a dimensionless coefficient $A_{1/5}$ that increases monotonically with the scaled interaction range $\mathcal{L}/\sqrt{\mathcal{S}}$.
Details on the composite fermion results for the $\nu=1/5$ state are presented in App.~\ref{app:nu_15}.
Having established the same scaling law for both filling factors, we can directly compare the cavity-induced gap enhancement under identical physical sample and cavity parameters.

To make the dependence on the filling factor explicit, we rewrite Eq.~(\ref{eqn:scalingLaw}) in terms of the physical cavity parameters using the definition of $\xi$ together with the relation
$
N=\nu N_{\phi}=\nu\mathcal{S}/(2\pi\ell^2),
$
obtaining
\begin{align}\label{eqn:gap_v_nu}
    \delta \Delta^{(\nu)}_{\rm ch}
    &=
    \left[
        \frac{
            3 e^4 \mathcal{S}^2
        }{
            32 \pi^2  m^2 \hbar\omega_{\rm cav}^5
        }
    \right]
    \mathcal{G}_{\rm E}^4 \;
    \nu^2
    A_{\nu} \left( \mathcal{L} / \sqrt{\mathcal{S}} \right)
    \, .
\end{align}

Equation~(\ref{eqn:gap_v_nu}) separates the dependence on the fractional quantum Hall state from that on the cavity parameters.
All filling-factor dependence is contained in the dimensionless factor
$\nu^2A_{\nu}(\mathcal{L}/\sqrt{\mathcal{S}})$,
whereas the remaining prefactor depends only on the cavity characteristics and the physical size of the sample.
This form therefore provides a direct framework for comparing the cavity-induced enhancement of the charge gap at different filling factors under identical experimental conditions,
where the surface area $\mathcal{S}$ is kept constant.

\begin{figure}[t!]
    \centering
    \begin{overpic}[width=\columnwidth]{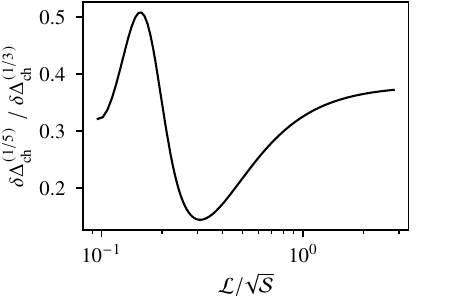}
        \put(20,60){\large (a)}
    \end{overpic}
    \vspace{1.35em}
    \begin{overpic}[width=\columnwidth]{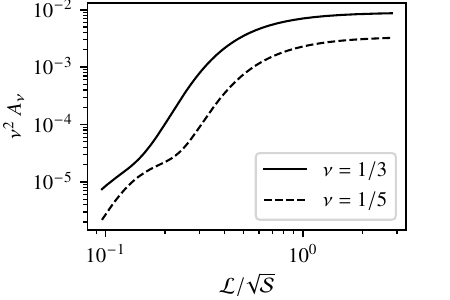}
        \put(21,60){\large (b)}
    \end{overpic}
\caption{
(a) Ratio of the cavity-induced charge-gap enhancements for the Laughlin $\nu=1/5$ and $\nu=1/3$ states as a function of the scaled interaction range $\mathcal{L}/\sqrt{\mathcal{S}}$, as obtained from Eq.~(\ref{eqn:gap_v_nu}).
The coefficients $A_{\nu}(\mathcal{L}/\sqrt{\mathcal{S}})$ are extracted from finite-size scaling analyses using $N=12,16,20,25,35$, and $50$ electrons.
The corresponding data for the $\nu=1/5$ state are presented in App.~\ref{app:nu_15}.
(b) Dependence of the dimensionless enhancement factor $\nu^2A_{\nu}(\mathcal{L}/\sqrt{\mathcal{S}})$ on the scaled interaction range for the $\nu=1/3$ and $\nu=1/5$ Laughlin states.
}

    \label{fig:delta13_delta15}
\end{figure}

In Fig.~\ref{fig:delta13_delta15}, we compare the cavity-induced enhancement of the charge gap for the Laughlin states at filling factors $\nu=1/3$ and $\nu=1/5$.
The enhancement is positive in both cases, demonstrating that the cavity-mediated interaction increases the incompressibility of both fractional quantum Hall states.
Its magnitude, however, differs substantially between the two filling factors.
As shown by Eq.~(\ref{eqn:gap_v_nu}), this dependence is entirely captured by the dimensionless factor
$\nu^2A_{\nu}(\mathcal{L}/\sqrt{\mathcal{S}})$.
In particular, the enhancement of the charge gap is significantly smaller for the $\nu=1/5$ state than for the $\nu=1/3$ state under otherwise identical physical conditions.
Furthermore, the ratio of the two gap enhancements depends sensitively on the scaled interaction range $\mathcal{L}/\sqrt{\mathcal{S}}$.

Our analysis provides quantitative predictions for the cavity-induced enhancement of the charge gap throughout the Laughlin sequence $\nu=1/(2p+1)$.
For these states, Eq.~(\ref{eqn:gap_v_nu}) shows that the dependence on the filling factor is entirely contained in the dimensionless combination $\nu^2A_{\nu}$.
We now exploit particle-hole symmetry to infer the corresponding behavior at the conjugate filling factors $(1-\nu)$.

Neglecting spin and Landau-level mixing, particle-hole symmetry within a single spin-polarized Landau level relates the cavity-induced correction to the charge gap at filling factor $\nu$ to that at filling factor $(1-\nu)$.
For fixed physical parameters, this symmetry implies
$
    \delta \Delta_{\rm ch}^{(\nu)}
    =
    \delta \Delta_{\rm ch}^{(1-\nu)} \, .
$
Using Eq.~(\ref{eqn:gap_v_nu}) for the two conjugate filling factors while keeping $\mathcal{L}/\sqrt{\mathcal{S}}$ fixed, we obtain
\begin{equation}
    \nu^2 A_{\nu}
    =
    (1-\nu)^2 A_{1-\nu} \, .
\end{equation}
Consequently, the cavity-induced enhancement is identical for particle-hole-conjugate states, for example $\nu=1/5$ and $\nu=4/5$.

This relation implies that, for identical cavity and sample parameters, the enhancement of the charge gap at $\nu=4/5$ is smaller than that at $\nu=1/3$.
Neglecting spin mixing, the state at $\nu=4/3$ can be viewed as an inert filled Landau level together with an active spin-flipped $\nu=1/3$ component of the lowest Landau level.
The cavity-induced enhancement of the charge gap is therefore expected to be larger at $\nu=4/3$ than at $\nu=4/5$.
This prediction is qualitatively consistent with the experimental observations reported in Ref.~\cite{Enkner}.

\subsection{Cavity-induced modification of the neutral spectrum}

We now investigate how the cavity-mediated interaction
modifies the neutral excitation spectrum of the Laughlin state at $\nu=1/3$.
For a composite-fermion exciton state $\psi^{\rm exc}_{k}$,
the cavity-induced correction to the neutral excitation gap is defined as
\begin{equation}
    \delta\Delta_k
    =
    \bra{\psi^{\rm exc}_{k}}
    \hat{V}_{\rm cav}^{\mathcal{L}}
    \ket{\psi^{\rm exc}_{k}}
    -
    \bra{\psi_{\rm GS}}
    \hat{V}_{\rm cav}^{\mathcal{L}}
    \ket{\psi_{\rm GS}} \, ,
\end{equation}
where $k$ denotes the wave vector
of the composite-fermion exciton.

A particularly important regime
is the large-$k$ limit of the neutral excitation spectrum.
Within the composite-fermion picture,
the large-$k$ neutral excitation corresponds to a quasielectron
and a quasihole separated by a large distance.
Its energy is therefore closely related
to the transport gap~\cite{bonesteel1995composite}.
For finite systems, however, it generally differs from the charge gap
because of the residual interaction between the quasielectron and quasihole.
Since this interaction depends on the range of the cavity-mediated interaction,
the cavity-induced corrections to the charge gap
and to the large-$k$ neutral gap are, in general, 
not expected to coincide for finite values of the cutoff length $\mathcal{L}$.

\begin{figure}[t!]
    \centering
    \includegraphics[width=\columnwidth]{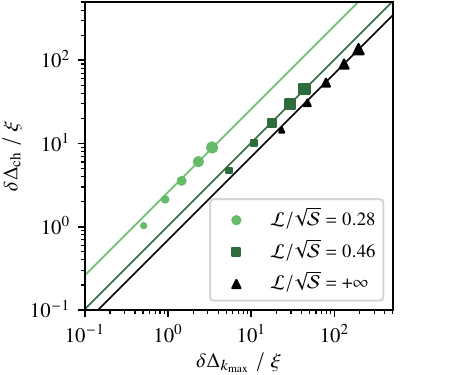}
\caption{
    Scatter plot of the cavity-induced correction to the charge gap $\delta \Delta_{\rm ch}$
    versus that of the neutral excitation at the largest
    quasielectron-quasihole separation $\delta \Delta_{k_{\rm max}}$ for the $\nu=1/3$ Laughlin state.
    The data are obtained for $N=12,16,20,25$, and $30$ electrons,
    with larger markers corresponding to larger number of electrons,
    and for different values of the scaled interaction range $\mathcal{L}/\sqrt{\mathcal{S}}$.
    The solid lines are linear fits,
    demonstrating that the two quantities are nearly proportional
    over the range of parameters considered,
    following the same scaling law as $N$ increases.
}
    \label{fig:scatterPlot_charge_highq}
\end{figure}

In Fig.~\ref{fig:scatterPlot_charge_highq},
we compare the cavity-induced correction to the charge gap $\delta \Delta_{\rm ch}$
with that of the composite-fermion exciton
at the largest quasielectron-quasihole separation $\delta \Delta_{k_{\rm max}}$.
For each value of the interaction range,
increasing the number of electrons moves the data points
toward larger values of both quantities.
The approximately linear relation between them
therefore shows that $\delta\Delta_{\rm ch}$ and $\delta\Delta_{k_{\rm max}}$
follow the same finite-size scaling as $N$ is varied.
The two quantities are nearly proportional
over the range of system sizes and interaction ranges considered,
with a proportionality coefficient of order unity.
The weak dependence of this coefficient on the interaction range
is consistent with the residual quasielectron-quasihole interaction discussed above.
These results demonstrate that the characteristic energy scale
governing the cavity-induced enhancement of the charge gap
also controls the high-$k$ neutral excitation.

\begin{figure}[t!]
    \centering
    \includegraphics[width=\columnwidth]{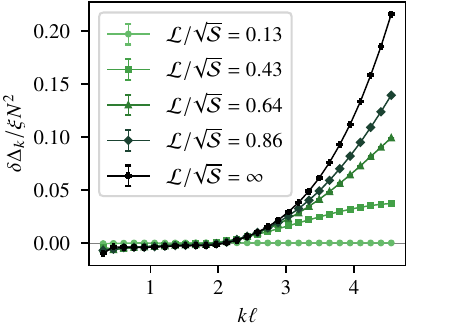}
    \caption{
        Cavity-induced correction to the neutral excitation spectrum
        of the Laughlin $\nu=1/3$ state for $N=30$ electrons.
        The correction to the neutral excitation gap, $\delta\Delta_k$,
        is shown in units of the characteristic interaction energy scale $\xi N^2$,
        with $\xi$ defined in Eq.~(\ref{eqn:xi}),
        as a function of the neutral-mode wave vector $k$
        for several values of the scaled interaction range $\mathcal{L}/\sqrt{\mathcal{S}}$,
        where $\mathcal{S}$ denotes the surface area.
}
    \label{fig:deltaDeltak_vs_calL}
\end{figure}

Figure~\ref{fig:deltaDeltak_vs_calL} shows the cavity-induced shift
of the neutral excitation branch for the Laughlin $\nu=1/3$ state with $N=30$ electrons.
For all interaction ranges considered, the correction depends on the wave vector $k$.
Both its magnitude and its momentum dependence are controlled by the range of the cavity-mediated interaction.
For the shortest cutoff length shown, the overall shift is much smaller than for longer interaction ranges and therefore appears nearly flat on the scale of the figure.
As the cutoff length increases, the momentum dependence becomes progressively more pronounced.

At low and intermediate momenta, the cavity-induced correction remains comparatively small.
By contrast, for sufficiently long interaction ranges it increases rapidly at large $k$, with the largest positive shift obtained in the untruncated limit
$\mathcal{L}/\sqrt{\mathcal{S}}\rightarrow+\infty$.
This behavior is consistent with the interpretation of the high-$k$ composite-fermion exciton as a widely separated quasielectron-quasihole pair, whose energy is governed by the same characteristic scale that controls the cavity-induced enhancement of the charge gap.
The cavity-mediated interaction therefore does not simply produce an overall upward shift of the neutral excitation spectrum.
Instead, it reshapes the dispersion in a manner that depends on both the interaction range and the wave vector, with the strongest enhancement occurring in the large-$k$ region of the magnetoroton branch.

\begin{figure*}[t!]
    \centering
    \resizebox{0.99\textwidth}{!}{%
        \includegraphics[width=\columnwidth]{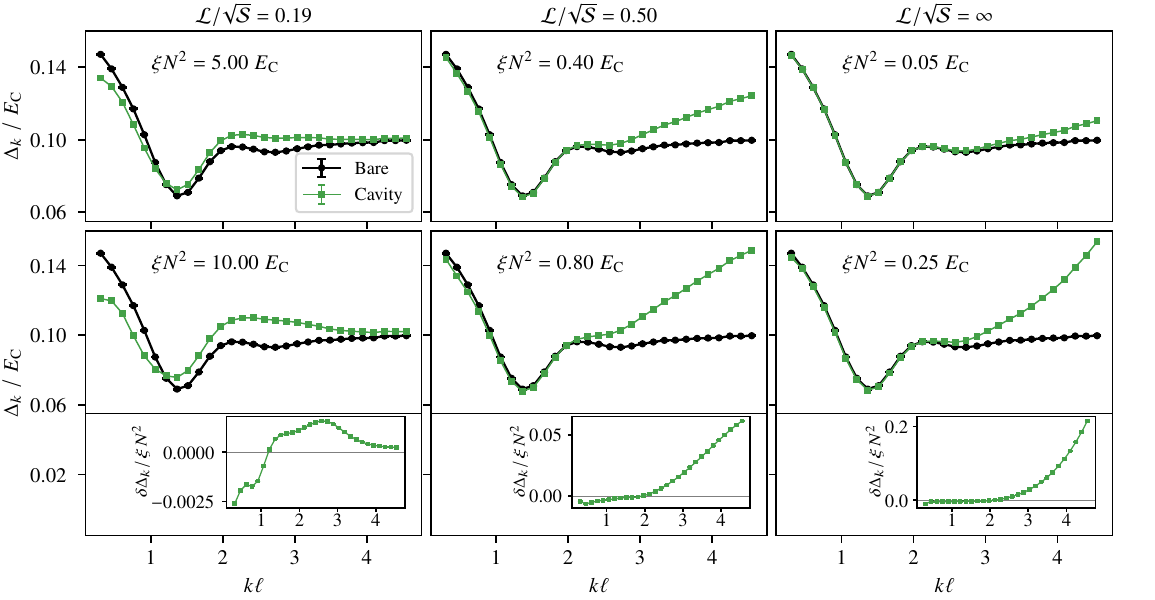}
    }
    \caption{
        Cavity-modified neutral excitation spectrum of the Laughlin state at $\nu=1/3$ for $N=30$ electrons.
        The neutral excitation gap $\Delta_k$, 
        reported in units of the Coulomb energy scale 
        $E_{\mathrm{C}} \equiv e^2 / (4 \pi \epsilon \ell)$,
        is plotted as a function of the dimensionless neutral-mode wave vector $k\ell$.
        Each column corresponds to a different scaled cutoff length 
        $\mathcal{L}/\sqrt{\mathcal{S}}$, as indicated above the panels,
        while the two rows correspond to different values of the cavity-induced interaction strength $\xi N^2$,
        with $\xi$ defined in Eq.~(\ref{eqn:xi}),
        as indicated in each panel.
        Black dots represent the bare Coulomb spectrum, 
        whereas green squares represent the cavity-modified spectrum.
        The insets show the cavity-induced gap shifts $\delta \Delta_k$,
        reported in units of $\xi N^2$.
        We note that the shift $\delta \Delta_k/ \xi N^2$
        does not depend on the scaled coupling $\xi N^2$.
    }
    \label{fig:magRotGMP_cavAndCoulomb_6pannel}
\end{figure*}

We now use the momentum-dependent corrections obtained above to reconstruct the cavity-modified magnetoroton dispersion.
Specifically, we treat the cavity-mediated interaction as a correction to the bare Coulomb spectrum by adding the cavity-induced shift $\delta\Delta_k$ to the neutral excitation energy $\Delta_k^{\rm C}$ computed for the Coulomb interaction alone,
\begin{equation}
    \Delta_k
    =
    \Delta_k^{\rm C}
    +
    \delta\Delta_k \, .
\end{equation}
This procedure yields the neutral excitation spectrum in the presence of the cavity-mediated interaction while retaining the Coulomb magnetoroton branch as the reference dispersion.

Figure~\ref{fig:magRotGMP_cavAndCoulomb_6pannel}
shows the resulting neutral excitation spectra
for three representative interaction ranges.
Each column corresponds to a fixed interaction range, 
set by the value of $\mathcal{L}/\sqrt{\mathcal{S}}$ indicated at the top of the column.
For each interaction range,
two representative values of the cavity-induced interaction strength are shown,
with the corresponding value of $\xi N^2$ indicated inside each panel.
In each case, the black curve represents the bare Coulomb magnetoroton branch,
while the green curve includes the cavity-induced correction.
The cavity-mediated interaction does not simply produce a rigid upward shift of the spectrum.
Instead, it modifies the magnetoroton dispersion in a momentum-dependent manner,
with the magnitude and shape of the distortion determined by the interaction range.

As shown in Fig.~\ref{fig:magRotGMP_cavAndCoulomb_6pannel},
the cavity-mediated interaction modifies the magnetoroton dispersion
in a manner that depends sensitively on the interaction range.
In particular, the roton minimum can either increase or decrease.
For interaction ranges up to approximately one third of the system size, however,
we consistently observe an enhancement of the roton gap,
as can be seen in the first column of
Fig.~\ref{fig:magRotGMP_cavAndCoulomb_6pannel}.

We note that our earlier analysis based on the Girvin-MacDonald-Platzman (GMP) formalism~\cite{Enkner}
predicted a roton-gap enhancement scaling as $\sim \mathcal{L}^4$.
That analysis considered a different thermodynamic scaling,
in which the $N \to \infty$ limit was taken first,
before studying the dependence on $\mathcal{L}$.
In the present work we keep the ratio
$\mathcal{L}/\sqrt{\mathcal{S}}$ fixed.
In our CF simulations,
we have noticed a consistent enhancement of the roton gap
for $\mathcal{L}/\sqrt{\mathcal{S}}\lesssim0.3$
and electron numbers up to $N=30$;
see, for example, the left column in Fig.~\ref{fig:magRotGMP_cavAndCoulomb_6pannel}.
This result therefore remains qualitatively consistent with the earlier GMP prediction,
while the present calculations further show that the evolution of the roton minimum
is not universal, but depends sensitively on the interaction range.

By contrast, the transport gap,
corresponding to the large-$k$ limit of the neutral excitation spectrum,
is enhanced for all interaction ranges considered.
This robust behavior reflects its close connection to the charged excitation gap,
as discussed above, and shows that the cavity-induced enhancement of the incompressibility
is considerably more robust than the behavior of the roton minimum.

In App.~\ref{app:ED_benchmarks}, we benchmark the neutral excitation spectra obtained within the composite-fermion approach against exact diagonalization.
Since only relatively small system sizes are accessible to exact diagonalization, the interaction strength $\xi$ is increased so as to keep the scaled coupling $\xi N^2$ fixed.
Even under these conditions, the cavity-induced modification of the neutral excitation spectrum remains of the order of $A_{\nu}\,\xi N^2$.
The comparison shows excellent agreement between the two approaches, providing strong support for the validity of the composite-fermion description employed throughout this work.

Importantly, using the nominal experimental parameters reported in Ref.~\cite{Enkner}, namely
$
    f_{\rm cav} = 0.1 \,{\rm THz}, \;
    B = 6.4 \,{\rm T}, \;
    N = 10^7, \;
    \mathcal{G}_{\rm E} = 4 \times 10^8 \,{\rm Vm^{-2}}, \;
    \epsilon_r = 13, \;
    m = 0.067 m_{\rm e},
$
we obtain a scaled interaction strength
$
    \xi N^2 \simeq 0.05\,E_{\rm C}.
$
Moreover, within the present minimal model, the characteristic interaction energy scales as
$\xi\propto\mathcal{G}_{\rm E}^{\,4}$.
Consequently, even moderate increases of the cavity-mode spatial gradient can produce substantially larger cavity-induced modifications of the fractional quantum Hall excitation spectrum.

\section{\label{sec:conclusions}
Conclusions and perspectives
}

In this work, we have investigated how cavity-mediated electron-electron interactions modify the excitation spectrum of fractional quantum Hall states within the composite-fermion framework. Compared with our previous analysis based on the Girvin-MacDonald-Platzman formalism~\cite{Enkner}, the composite-fermion approach provides a quantitatively more accurate description of both charged and neutral excitations and enables the calculation of the cavity-induced modification of the charge excitation gap, the quantity directly relevant to transport experiments.

Our analysis reveals a distinctive finite-size scaling law for the cavity-induced enhancement of the charge gap. Owing to the long-range character of the cavity-mediated interaction, the gap enhancement scales quadratically with the number of electrons, while its dependence on the interaction range is entirely captured by the dimensionless ratio $\mathcal{L}/\sqrt{\mathcal{S}}$. This scaling enables a well-defined thermodynamic extrapolation and provides quantitative predictions for the cavity-induced modification of the charge gap throughout the Laughlin sequence. Furthermore, exploiting particle-hole symmetry, we extended these predictions to the conjugate filling factors.

We have also investigated the cavity-induced modification of the neutral excitation spectrum. While the cavity-mediated interaction reshapes the magnetoroton branch in a momentum- and range-dependent manner, the large-$k$ neutral excitation is found to be governed by the same characteristic energy scale that controls the charged excitation gap. Consequently, the transport gap is enhanced for all interaction ranges considered, whereas the roton minimum can either increase or decrease depending on the spatial extent of the cavity-mediated interaction. Benchmarks against exact diagonalization for the system sizes accessible numerically show excellent agreement with the composite-fermion calculations, providing strong support for the validity of the present approach.

The present work has focused on orbital effects within the lowest Landau level and on a minimal cavity-mediated interaction characterized by a spatially uniform field gradient with a finite interaction range. Several natural extensions emerge from this study. A first direction is to investigate more realistic cavity geometries and electromagnetic mode profiles, which can generate a broader class of effective interactions and enable cavity pseudopotential engineering of fractional quantum Hall states. Another important extension is the inclusion of spin degrees of freedom. Cavity-induced modifications of the electronic $g$ factor may alter the competition between spin-polarized, partially spin-polarized, and spin-unpolarized fractional quantum Hall phases, particularly in regimes where these states are nearly degenerate. Finally, extending the present framework to higher Landau levels may provide new opportunities to manipulate the competition between correlated phases, including even-denominator fractional quantum Hall states and composite Fermi-liquid phases, through cavity-engineered electron-electron interactions.

\begin{acknowledgments}
D.B. thanks Kishore Iyer for numerous discussions on FQH physics.
We thank Lorenzo Graziotto, Josefine Enkner, Ethan Koskas, Giacomo Scalari
and Jérôme Faist for many fruitful discussions.
We acknowledge financial support from the French agency ANR 
through the project CaVdW (ANR-21-CE30-0056-0) and the project TEASER (ANR-24-CE24-4830). The Flatiron
Institute is a division of the Simons Foundation. 
\end{acknowledgments}

\appendix

\section{\label{app:theory}
    Theoretical Framework
}

In this appendix, we derive the effective cavity-mediated electron-electron interaction underlying the results presented in the main text. We begin with the Hamiltonian describing the quantum Hall system in the absence of light-matter coupling and then introduce its interaction with the cavity field. Projecting onto the lowest Landau level and adiabatically eliminating the photonic degrees of freedom in the off-resonant regime yields an effective electronic Hamiltonian containing a cavity-mediated electron-electron interaction. Finally, we derive a representative real-space pair potential associated with this effective two-body interaction.

\subsection{\label{app:QH_Hamiltonian}
    Quantum Hall Hamiltonian
}
We consider a two-dimensional electron gas subject to a perpendicular magnetic field $B$.
In the absence of interactions, the single-particle Hamiltonian is
$
\hat{h}_{\rm kin} = \hat{\pi}_{\mu}\hat{\pi}_{\mu}/2m,
$
where
$
\hat{\pi}_{\mu} = \hat{p}_{\mu} + eA_{\mu}
$
is the kinetic momentum,
$m$ is the electronic band mass,
$-e$ is the electron charge,
$\hat{p}_{\mu}$ is the canonical momentum conjugate to the position operator $\hat{r}_{\mu}$,
and $A_{\mu}$ is a vector potential generating the uniform magnetic field
$
B = \epsilon_{\mu\nu}\,\partial_{\mu}A_{\nu}.
$
Greek indices label the two spatial directions $x$ and $y$, and repeated indices are implicitly summed over.

It is convenient to decompose the position operator as
$
    \hat{r}_{\mu} = \hat{R}_{\mu} + \hat{\eta}_{\mu} \, ,
$
where $\hat{R}_{\mu}$ and $\hat{\eta}_{\mu}$ denote the guiding-center and cyclotron-radius coordinates, respectively.
The cyclotron-radius coordinates are related to the kinetic momentum through
\begin{equation}
\hat{\eta}_{\mu} = (\ell^2/\hbar)\epsilon_{\mu\nu}\hat{\pi}_{\nu} \, ,
\end{equation}
where
$
\ell \equiv \sqrt{\hbar/eB}
$
is the magnetic length.
These operators satisfy the commutation relations
$
[\hat{R}_{\mu}, \hat{R}_{\nu}] = \img \epsilon_{\mu\nu}\ell^2,
$
$
[\hat{\eta}_{\mu}, \hat{\eta}_{\nu}] = -\img \epsilon_{\mu\nu}\ell^2,
$
and
$
[\hat{R}_{\mu}, \hat{\eta}_{\nu}] = 0.
$

We define the cyclotron and guiding-center ladder operators as
\begin{align}
\hat{a}
&=
\frac{1}{\sqrt{2}\ell}
\left(
\hat{\eta}_x - \img \hat{\eta}_y
\right)\, ,
\\
\hat{b}
&=
\frac{1}{\sqrt{2}\ell}
\left(
\hat{R}_x + \img \hat{R}_y
\right)\, .
\end{align}
They satisfy the commutation relations
$
[\hat{a},\hat{a}^{\dagger}] 
=
[\hat{b},\hat{b}^{\dagger}] = 1,
$
and
$
[\hat{a},\hat{b}] = [\hat{a},\hat{b}^{\dagger}] = 0.
$

With these definitions, the kinetic Hamiltonian takes the form
$
\hat{h}_{\rm kin}
=
\hbar\omega
\left(
\hat{a}^{\dagger}\hat{a}
+
\frac12
\right),
$
revealing a spectrum of equally spaced, dispersionless energy bands known as Landau levels. The spacing between adjacent Landau levels is given by the cyclotron energy $\hbar\omega$, where
$
\omega \equiv eB/m
$
is the cyclotron frequency.

Each Landau level is macroscopically degenerate, with a degeneracy determined by the number of magnetic flux quanta threading the system. For a sample of area $\mathcal{S}$, the degeneracy is
$
N_{\phi} = \phi/\phi_0 = \mathcal{S}/(2\pi \ell^2),
$
where
$
\phi_0 = h/e
$
is the magnetic flux quantum.
The many-body problem is naturally characterized by the filling factor
$
\nu = N/N_{\phi},
$
which measures the ratio between the number of electrons $N$ and the Landau-level degeneracy.
Equivalently, the filling factor can be written as
$
\nu = n_{\rm e} 2\pi \ell^2,
$
where $n_{\rm e}$ is the electron density.

The many-body electronic Hamiltonian is
\begin{equation}
\hat{H}_{\rm el}
=
\sum_i \hat{h}_{\rm kin}^{(i)}
+
\hat{V}_{\rm C} \, , 
\end{equation}
with
\begin{equation}
\hat{V}_{\rm C}
=
\frac{e^2}{4\pi\epsilon}
\sum_{i<j}
\frac{1}{
\left|
\hat{\bm{r}}^{(i)}
-
\hat{\bm{r}}^{(j)}
\right|
} \, ,
\end{equation}
describing the electron-electron Coulomb interaction, and
$
\epsilon=\epsilon_0\epsilon_r
$
the electric permittivity of the host material.

In the regime of strong magnetic fields, electrons partially occupy only the lowest Landau level (LLL), while the cyclotron energy exceeds the characteristic Coulomb interaction scale,
\begin{equation}
E_{\rm cyc}
\equiv
\hbar\omega
\gg
\frac{e^2}{4\pi\epsilon\ell}
\equiv
E_{\rm C}.
\end{equation}
Under these conditions, Landau-level mixing induced by the Coulomb interaction can be neglected. The low-energy physics is therefore entirely governed by the Coulomb interaction projected onto the LLL,
\begin{equation}
\hat{V}_{\rm C}^{\rm LLL}
\equiv
\hat{\mathcal{P}}_{\rm LLL}
\hat{V}_{\rm C}
\hat{\mathcal{P}}_{\rm LLL} \, ,
\end{equation}
where $\hat{\mathcal{P}}_{\rm LLL}$ denotes the projector onto the lowest Landau level. This projected Hamiltonian supports a rich variety of strongly correlated phases, whose understanding is rooted in the emergence of composite fermions, topological quasiparticles formed by binding electrons to quantum magnetic fluxes \cite{jainBook}.

Both electrons and composite fermions possess spin degrees of freedom. In the regime considered here, however, the large Zeeman splitting associated with the strong magnetic field effectively freezes the spin dynamics and renders the system fully spin polarized. Consequently, spin degrees of freedom are neglected throughout this work, and we focus exclusively on the orbital dynamics within a single spin sector. The inclusion of spin effects is left for future investigations.

\subsection{\label{app:QH_cav}
    Light Matter Coupling Hamiltonian
}
In what follows, we consider a quantum Hall system coupled to a single quantized, linearly polarized cavity mode, following Ref.~\cite{Enkner}.
While the cavity field is three-dimensional, the electrons are confined to a plane, so that only the in-plane components of the vector potential couple to the electronic degrees of freedom. We therefore take the in-plane cavity vector-potential operator to be
\begin{equation}
    \hat{A}^{(\rm cav)}_{\mu}
    =
    A_0\bigl(\delta_{\mu x}+f_{\mu}(\hat{r})\bigr)
    (\hat{\mathcal{A}}+\hat{\mathcal{A}}^{\dagger}) \, ,
\end{equation}
where $\hat{\mathcal{A}}^{\dagger}$ ($\hat{\mathcal{A}}$) creates (annihilates) a photon in the cavity mode. The parameter $A_0$ sets the overall field amplitude, while $f_{\mu}$ describes spatial variations of the cavity field around its uniform component.

We work in the Coulomb gauge,
$
\bm{\nabla}_{\rm 3D}\cdot\hat{\bm{A}}^{(\rm cav)} = 0.
$
Assuming the out-of-plane component of the cavity field to be spatially uniform, this condition reduces to the two-dimensional transversality condition
$
\partial_{\mu}f_{\mu}=0.
$

The light-matter coupling Hamiltonian is obtained through the minimal-coupling substitution
$
\hat{\pi}_{\mu}
\rightarrow
\hat{\pi}_{\mu}
+
e\hat{A}^{(\rm cav)}_{\mu}
$
in $\hat{H}_{\rm el}$.
Expanding the kinetic-energy term and using the Coulomb-gauge condition, which is equivalent to
$
[\hat{\pi}_{\mu},\hat{A}^{(\rm cav)}_{\mu}] = 0,
$
the total Hamiltonian can be written as
$
\hat{H}
=
\hat{H}_{\rm el}
+
\hat{H}_{\rm ph}
+
\hat{H}_{\rm para}
+
\hat{H}_{\rm dia},
$
with
\begin{align}
    \hat{H}_{\rm ph}
    &=
    \hbar \omega_{\rm cav}\,
    \hat{\mathcal{A}}^{\dagger}\hat{\mathcal{A}}
    +
    g_{\rm D}N
    \left(
        \hat{\mathcal{A}}
        +
        \hat{\mathcal{A}}^{\dagger}
    \right)^2 ,
    \\
    \nonumber
    \hat{H}_{\rm para}
    &=
    g_{\rm P}
    \left(
        \hat{\mathcal{A}}
        +
        \hat{\mathcal{A}}^{\dagger}
    \right)
    \sum_i
    (\ell/\hbar)\hat{\pi}_{x}^{(i)}
    \\
    &+
    g_{\rm P}
    \left(
        \hat{\mathcal{A}}
        +
        \hat{\mathcal{A}}^{\dagger}
    \right)
    \sum_i
    (\ell/\hbar)
    f_{\mu}\!\left(\hat{r}^{(i)}\right)
    \hat{\pi}_{\mu}^{(i)},
    \\
    \nonumber
    \hat{H}_{\rm dia}
    &=
    2g_{\rm D}
    \left(
        \hat{\mathcal{A}}
        +
        \hat{\mathcal{A}}^{\dagger}
    \right)^2
    \sum_i
    f_x\!\left(\hat{r}^{(i)}\right)
    \\
    &+
    g_{\rm D}
    \left(
        \hat{\mathcal{A}}
        +
        \hat{\mathcal{A}}^{\dagger}
    \right)^2
    \sum_i
    f_{\mu}\!\left(\hat{r}^{(i)}\right)
    f_{\mu}\!\left(\hat{r}^{(i)}\right).
\end{align}

We have defined the paramagnetic and diamagnetic coupling strengths as
\begin{align}
    g_{\rm P}
    &\equiv
    \frac{eA_0\hbar}{m\ell},
    \\
    g_{\rm D}
    &\equiv
    \frac{e^2A_0^2}{2m},
\end{align}
and denoted by $\omega_{\rm cav}$ the cavity angular frequency, which determines the vacuum electric-field amplitude through the relation:
\begin{equation}
E_{\rm vac}
=
\omega_{\rm cav}A_0\, .
\end{equation}

The homogeneous diamagnetic contribution proportional to $A_0^2$ and to the electron number has been absorbed into a redefinition of the photonic quadratures through a Bogoliubov transformation, introducing a squeezing parameter \cite{CiutiPRB21}. In the parameter regime relevant to the present work, this parameter remains very close to unity, implying that the resulting renormalization of both the light-matter coupling strengths and the cavity frequency is negligible. We therefore neglect this contribution and retain the same notation for the coupling constants and cavity frequency throughout.

In the absence of spatial inhomogeneity,
the light–matter coupling generates only vertical inter-Landau-level
transitions governed by cyclotron coordinates,
in accordance with Kohn's theorem \cite{kohnsTheorem, rokajWeakened}.
As a result, no intra-Landau-level dynamics arise in this limit.
To induce nontrivial dynamics within a single Landau level,
it is necessary to violate the conditions underlying Kohn's theorem,
namely the assumption of a spatially uniform electromagnetic field.
This is achieved by introducing a finite spatial variation
of the cavity mode, encoded in a nonzero $f_{\mu}(\hat r)$.
As a minimal model, we consider a linear spatial dependence,
\begin{equation}
    f_{\mu}(\hat r)=\mathcal{G}\hat r_x\delta_{\mu y} \, .
\end{equation}
We note that the gradient parameter $\mathcal{G}$ introduced here 
has dimensions $\,{\rm L}^{-1}$ and can be related to the 
vector potential gradient parameter $\mathcal{G}_{\rm A}$
and the experimentally relevant $\mathcal{G}_{\rm E}$
from Ref.~\cite{Enkner} as
\begin{align}
    \mathcal{G}_{\rm A}
    &\equiv
    A_0 \mathcal{G} \, ,
    \\
    \mathcal{G}_{\rm E}
    &\equiv
    E_{\rm vac} \mathcal{G} = A_0 \omega_{\rm cav} \mathcal{G}  \, ,
\end{align}
as defined in the main text.

The paramagnetic and diamagnetic interactions reduce to the following expressions:
\begin{align}
    \hat H_{\rm para}
    &=
    \frac{g_{\rm P}\mathcal G}{\ell}
    \left(
        \hat{\mathcal A}
        +
        \hat{\mathcal A}^{\dagger}
    \right)
    \sum_i
    \left[
        \left(\hat\eta_x^{(i)}\right)^2
        +
        \hat R_x^{(i)}\hat\eta_x^{(i)}
        -
        \ell\,\hat\eta_y^{(i)}
    \right],
    \\
    \hat H_{\rm dia}
    &=
    g_{\rm D}\mathcal G^2
    \left(
        \hat{\mathcal A}
        +
        \hat{\mathcal A}^{\dagger}
    \right)^2
    \sum_i
    \left[
        \left(\hat\eta_x^{(i)}\right)^2
        +
        \left(\hat R_x^{(i)}\right)^2
        +
        2\hat R_x^{(i)}\hat\eta_x^{(i)}
    \right].
\end{align}

The next step is to project the Hamiltonian onto the lowest Landau level (LLL). The guiding-center coordinates are unaffected by the projection. Terms linear in $\hat{\eta}_\mu$ vanish because $\hat{\eta}_\mu$ connects different Landau levels and therefore has no matrix elements within the LLL. To evaluate the projection of $\hat{\eta}_x^2$, we express it in terms of the cyclotron ladder operators and retain only the diagonal contribution. This yields
\begin{equation}
    \mathcal P_{\rm LLL}
    \hat{\eta}_x^2
    \mathcal P_{\rm LLL}
    =
    \frac{\ell^2}{2}.
\end{equation}

The projected light-matter interactions then take the form
\begin{align}
    \hat H_{\rm para}^{\rm LLL}
    &=
    \frac{
        g_{\rm P}\mathcal G \ell N
    }{2}
    \left(
        \hat{\mathcal A}
        +
        \hat{\mathcal A}^{\dagger}
    \right),
    \\
    \hat H_{\rm dia}^{\rm LLL}
    &=
    g_{\rm D}\mathcal G^2
    \left(
        \hat{\mathcal A}
        +
        \hat{\mathcal A}^{\dagger}
    \right)^2
    \left(
        \frac{\ell^2 N}{2}
        +
        \sum_i
        \left(\hat R_x^{(i)}\right)^2
    \right).
\end{align}

After projection onto the LLL, the paramagnetic interaction is purely off-diagonal in photon number and does not involve guiding-center operators. Consequently, virtual processes generated by this term produce only a constant energy shift, which we disregard. By contrast, the diamagnetic interaction contains both diagonal and off-diagonal contributions involving the guiding-center degrees of freedom.
We note that beyond the present linear-gradient approximation, higher-order spatial variations of the cavity field can also generate paramagnetic couplings involving guiding-center operators.
In addition, after projection, the diamagnetic interaction contains a contribution $\propto N$,
which can be treated by the same Bogoliubov transformation discussed above.

We decompose the Hamiltonian as
\begin{equation}
\hat H
= 
\hat H_0
+
\hat V_{\rm d}
+
\hat V_{\rm od},
\end{equation}
where
\begin{equation}
\hat H_0
=
\hat V_{\rm C}^{\rm LLL}
+
\hbar\omega_{\rm cav}
\hat{\mathcal A}^{\dagger}\hat{\mathcal A},
\end{equation}
\begin{equation}
\hat V_{\rm d}
=
g_{\rm D}\mathcal G^2
\left(
1
+
2\hat{\mathcal A}^{\dagger}\hat{\mathcal A}
\right)
\hat{\mathcal{R}},
\end{equation}
\begin{equation}
\hat V_{\rm od}
=
g_{\rm D}\mathcal G^2
\left[
\hat{\mathcal A}^2
+
\left(\hat{\mathcal A}^{\dagger}\right)^2
\right]
\hat{\mathcal{R}}.
\end{equation}

Here, $\hat H_0$ defines the unperturbed Hamiltonian, $\hat V_{\rm d}$ contains the terms conserving the photon number, and $\hat V_{\rm od}$ collects the terms responsible for the creation and destruction of photons. We have introduced the operator

\begin{equation}
\hat{\mathcal{R}}
= 
\sum_i
\left(\hat R_x^{(i)}\right)^2.
\end{equation}

To adiabatically eliminate the photonic degrees of freedom, we perform a Schrieffer--Wolff transformation~\cite{SW} corresponding to the unitary transformation

\begin{equation}
\tilde H
= 
e^{\hat S}
\hat H
e^{-\hat S}
=
\hat H
+
[\hat S,\hat H]
+
\frac{1}{2}
[\hat S,[\hat S,\hat H]]
+
\mathcal O(\hat S^3),
\end{equation}
where the anti-Hermitian generator satisfies $\hat S^\dagger=-\hat S$ and is chosen such that
\begin{equation}
[\hat S,\hat H_0]
= 
-\hat V_{\rm od}.
\end{equation}
This choice eliminates the off-diagonal photon-number coupling to second order.
The exact formal solution involves the eigenstates $\ket{n}_{\rm c}$ and eigenvalues $E_n$ of $\hat V_{\rm C}^{\rm LLL}$:
\begin{equation}
\hat S
=
g_{\rm D}\mathcal G^2
\sum_{nn'}
\frac{
    \mathcal{R}_{n'n}
}{
E_{n'}-E_n+2\hbar\omega_{\rm cav}
}
|n'\rangle_{\rm c}\,
{}_{\rm c}\mkern-2mu\langle n|
\left(\hat{\mathcal A}^{\dagger}\right)^2
-
{\rm h.c.}.
\end{equation}
Here, $\mathcal{R}_{n'n} \equiv {}_{\rm c}\mkern-2mu\langle n'| \hat{\mathcal{R}} |n\rangle_{\rm c}$.

This expression explicitly depends on the excitation spectrum of the projected Coulomb Hamiltonian and therefore generates a state-dependent effective interaction. Expanding the transformed Hamiltonian to second order in $g_{\rm D}\mathcal G^2$ and subsequently projecting onto the photonic vacuum yields the effective Hamiltonian:
\begin{widetext}
\begin{equation}
\left(\hat H_{\rm eff}\right)_{nn'}
\simeq
E_n \delta_{nn'}
+
g_{\rm D}\mathcal G^2 {\mathcal{R}}_{nn'}
-
(g_{\rm D}\mathcal G^2)^2
\sum_m
{\mathcal{R}}_{nm} {\mathcal{R}}_{mn'}
\left[
\frac{1}{E_m-E_n+2\hbar\omega_{\rm cav}}
+
\frac{1}{E_m-E_{n'}+2\hbar\omega_{\rm cav}}
\right].
\end{equation}
\end{widetext}

The second-order contribution contains energy denominators associated with virtual transitions to excited Coulomb eigenstates. In general, this leads to a state-dependent effective interaction, since the corresponding resolvent depends on the projected Coulomb spectrum. We now consider the off-resonant regime, in which the cavity photon energy is much larger than the Coulomb energy differences associated with intermediate states having appreciable matrix elements of $\hat{\mathcal R}$. In this limit, the energy denominators can be approximated by the cavity photon energy, yielding to leading order the effective Hamiltonian:
\begin{equation}
\label{eqn:effHam}
\hat H_{\rm eff}
=
\hat V_{\rm C}^{\rm LLL}
+
g_{\rm D}\mathcal G^2
\hat{\mathcal{R}}
-
\frac{
\left(
g_{\rm D}\mathcal G^2
\right)^2
}{
\hbar\omega_{\rm cav}
}
\hat{\mathcal{R}}^2.
\end{equation}

This approximation replaces the full state-dependent
Coulomb energy resolvent by a single
effective energy scale, thereby reducing the interaction to a two-body form.
It neglects the detailed distribution of excitation energies and should be
understood as an effective description.

\subsection{\label{app:CavMedInt}
    Cavity-mediated pair potential
}
The cavity-mediated interaction in Eq.~(\ref{eqn:effHam})
contains both one-body and two-body contributions.
In this subsection, we shall derive all the terms
and retain those that we deem relevant for the many-body physics studied here.
The single-body contribution yields a confinement potential,
whereas the two-body terms contribute to a long-range attractive interaction.

After dropping the terms proportional to the identity,
the relevant cavity-induced contributions to the effective Hamiltonian are
\begin{align}
    \hat{V}_{\rm 1b;\,cav} 
    &=
    \sum_i V_{\rm 1b;\,cav}\left( \hat{R}_x^{(i)}  \right)
    \, ,
    \\
    \hat{V}_{\rm 2b;\,cav}
    &=
    -
    \frac{2 \xi }{3}
    \sum_{i < j} \hat{X}^{(i)} \hat{X}^{(j)} \, ,
\end{align}
where 
\begin{equation}
    \hat{X}
    \equiv
    2 \hat{b}^{\dagger}\hat{b}
    +
    \hat{b}^2
    +
    \hat{b}^{\dagger 2} \, ,
\end{equation}
such that $\hat{R}_x^2 = \ell^2(1 + \hat{X})/2$.
The two-body interaction strength is characterized by the energy scale
\begin{align}
\label{eqnapp:xi}
    \xi
    &\equiv
    \frac{
        3 g_{\rm D}^2 (\mathcal G \ell)^4
    }{
        2\hbar\omega_{\rm cav}
    }
    =
    \frac{
        3 e^4 (\mathcal{G}_{\rm E} \ell)^4
    }{
        8 m^2 \hbar \omega_{\rm cav}^5
    }
    \, ,
\end{align}
as defined in the main text in Eq.~(\ref{eqn:xi}).
The numerical prefactor $2/3$ is introduced for later convenience.
The one-body potential is defined as:
\begin{equation}
    V_{\rm 1b;\,cav}(x)
    \equiv
    \left[
        \sqrt{
            \frac{4 \hbar \omega_{\rm cav} \xi}{ 3}
        }
        - 
        \frac{4\xi}{3} (N-1)
    \right] \, 
    \left(\frac{x}{\ell}\right)^2
    -
    \left[\frac{4\xi}{3}\right] \,
    \left(\frac{x}{\ell}\right)^4 \, .
\end{equation}

The one-body contribution $\hat{V}_{\rm 1b;\, cav}$
acts as a cavity-induced correction to the lateral guiding-center confinement
along the direction of the cavity field gradient.
In a physical Hall bar,
such a term is naturally interpreted as a renormalization of the external confining potential
already present in the sample.
Its primary effect is therefore expected to be a modification of the density profile
and of edge properties, rather than of the intrinsic bulk physics. This interpretation is further supported by the small magnitude of the induced potential.
For the nominal experimental values \cite{Enkner}, $\xi N^2 = 0.05 E_{\rm C}$ and $N = 10^7$,
the cavity-induced confinement is weak.
Taking the linear dimension of the sample to be $~\sqrt{N} \ell$~\footnote{
At fixed density, the surface area is proportional to $N$.
The linear dimension is hence proportional to $\sqrt{N}$.
},
we obtain, for $x / \ell < 1000$, $| V_{\rm 1b;,cav}(x) |< 0.0008 E_{\rm C}$.
This is more than one order of magnitude smaller than
$\delta \Delta_{\rm ch} \sim 0.005 E_{\mathrm{C}}$,
the enhancement of the many-body charge gap produced by the cavity-induced two-body interaction
reported in the main text. Since our focus here is the intrinsic bulk physics of the fractional quantum Hall liquid,
we therefore regard this one-body contribution as being absorbed into the external confinement
and omit it in what follows.
We henceforth retain only the cavity-induced two-body interaction
$\hat{V}_{\rm 2b;\,cav}$.

To analyze the effects of the cavity-mediated electron-electron interaction,
we recast it in a form that makes explicit its dependence on relative guiding-center coordinates.
Unlike the Coulomb interaction, the effective interaction obtained above
is not exactly translationally or rotationally invariant.
This originates from the fixed polarization of the cavity mode,
the directional character of the cavity field gradient,
and the residual dependence on the choice of origin.
Consequently, the full interaction contains anisotropic and weakly inhomogeneous components,
which in principle could couple to geometric deformations of the fractional quantum Hall liquid
or favor symmetry-broken phases such as stripe
or nematic states in related settings~\cite{LorenzoStripes, SambuddhaStripes}.
A systematic study of these effects is left for future work.

In the present work, however, we focus on the leading isotropic bulk component of the
cavity-induced interaction.
This approximation is justified by the fact that the effect addressed here is the
renormalization of the incompressibility gap of an otherwise homogeneous fractional quantum Hall liquid.
For this purpose, the dominant contribution is the rotationally averaged interaction
in the relative guiding-center coordinate, while anisotropic and center-of-mass-dependent
terms describe subleading corrections that primarily couple to collective geometric
or edge-like distortions.
We therefore replace the full cavity-induced two-body operator by its
translationally and rotationally invariant projection.
Operationally, this is implemented by retaining only matrix elements
that are diagonal in the center-of-mass and relative angular momenta,
and by discarding the residual dependence on the center-of-mass quantum number.
The resulting interaction is fully characterized by Haldane pseudopotentials~\cite{HaldanePseudoPotentials}.

To derive the corresponding Haldane pseudopotentials,
we first introduce the single-particle guiding-center angular-momentum eigenstates
\begin{equation}
\ket{m}
=
\frac{\left(\hat b^\dagger\right)^m}{\sqrt{m!}}
\ket{0} \, ,
\end{equation}
where $\ket{0}$ denotes the orbital annihilated by both $\hat a$ and $\hat b$.
A convenient two-particle basis in the LLL is given by
\begin{equation}
\ket{M,m}
=
\frac{\left(\hat b_{\rm com}^{\dagger}\right)^M}{\sqrt{M!}}
\frac{\left(\hat b_{\rm rel}^{\dagger}\right)^m}{\sqrt{m!}}
\ket{0;0},
\end{equation}
where
\begin{equation}
\hat b_{\rm com}
=
\frac{
\hat b^{(1)}
+
\hat b^{(2)}
}{\sqrt{2}},
\qquad
\hat b_{\rm rel}
=
\frac{
\hat b^{(1)}
-
\hat b^{(2)}
}{\sqrt{2}},
\end{equation}
and $\ket{0;0}\equiv\ket{0}_1\otimes\ket{0}_2$.
The quantum numbers $M$ and $m$ correspond to the center-of-mass and relative angular momentum, respectively.
For the fully spin-polarized fermionic system considered here,
the relative angular momentum $m$ must be odd in order to satisfy the Pauli exclusion principle.

The interaction is fully characterized by its matrix elements
\begin{equation}
\bra{M,m}
\hat{V}_{\rm 2b;\,cav}
\ket{M',m'} \, ,
\end{equation}
from which the many-body Hamiltonian can be reconstructed.
To extract the rotationally and translationally invariant component of the interaction,
we retain only matrix elements that are diagonal in both center-of-mass and relative angular momentum
and eliminate any dependence on the center-of-mass quantum number:
\begin{widetext}
\begin{equation}
\bra{M,m}
\hat{V}_{\rm 2b;\,cav}
\ket{M',m'}
\rightarrow
\bra{M,m}
\hat{V}_{\rm 2b;\,cav}^{{\rm r}}
\ket{M',m'}
=
\delta_{MM'}
\delta_{mm'}
\bra{0,m}
\hat{V}_{\rm 2b;\,cav}
\ket{0,m}
\equiv
\delta_{MM'}
\delta_{mm'}
v_m^{(\rm cav)}.
\end{equation}
\end{widetext}
The quantities $v_m^{(\rm cav)}$ are the cavity-induced Haldane pseudopotentials.
By construction, this projection enforces invariance under guiding-center translations and rotations:
\begin{equation}
\left[
\hat{V}_{\rm 2b;\,cav}^{\rm r} \, ,
\sum_i \hat R_\mu^{(i)}
\right]
=
\left[
\hat{V}_{\rm 2b;\,cav}^{\rm r} \, ,
\sum_i
\hat R_\nu^{(i)}
\hat R_\nu^{(i)}
\right]
=
0.
\end{equation}
Evaluating the matrix elements yields:
\begin{align}\label{eqn:cavity_plane_pps}
v_m^{(\rm cav)}
&=
-\frac{2\xi}{3}
\bra{0,m}
\hat X^{(1)}
\hat X^{(2)}
\ket{0,m}
=
-\xi\left(m^2-m\right) \, .
\end{align}
We refer the reader to App.~\ref{app:PseudoPotentials} for details on the derivation.

These pseudopotentials can be used to construct
a corresponding real-space representation of $\hat{V}_{\rm 2b;\,cav}^{{\rm r}}$.
In the LLL, the relative angular momentum $m$
is related to the characteristic interparticle separation through $r\sim\sqrt{m}\ell$.
For example, the Coulomb pseudopotentials decay as $V_m\sim m^{-1/2}$,
consistent with the real-space interaction $V_{\rm C}(r)\propto 1/r$ \cite{girvinBook}.
Motivated by this correspondence,
we adopt a polynomial ansatz for the effective pair potential
and determine its coefficients by fitting them to the pseudopotentials $v_m^{(\rm cav)}$.
This procedure yields the representative interaction
\begin{equation}
\label{eqnapp:harmonicsCavPlane}
V_{\rm cav}(r)
=
-\xi
\left(
\frac{(r/\ell)^4}{16}
-
(r/\ell)^2
+
2
\right).
\end{equation}

This representation makes it explicit that the magnitude of the interaction increases with interparticle separation,
while the overall negative sign implies that the interaction is attractive.

The increase at large distances originates from the idealized assumption of a cavity mode with a spatially uniform gradient $\mathcal G$. In a realistic device, this behavior is regularized by a finite characteristic length scale. We therefore introduce a cutoff length $\mathcal L$, which we associate with the spatial extent over which the cavity field exhibits appreciable gradients. To regularize the interaction, we multiply the pair potential by a Gaussian envelope,
\begin{equation}
\label{eqnapp:harmonicsCavPlaneCut}
V_{\rm cav}(r;\mathcal L)
=
-\xi
\left(
\frac{(r/\ell)^4}{16}
-
(r/\ell)^2
+
2
\right)
e^{-r^2/\mathcal L^2} \, ,
\end{equation}
as given in Eq.~(\ref{eqn:harmonicsCavPlaneCut}) in the main text.
Throughout this work, we have denoted this pair potential by $V_{\rm cav}(r;\mathcal L)$
and the corresponding LLL-projected interaction by $\hat V_{\rm cav}^{\mathcal L}$.

\section{Composite-fermion ansatz}
\label{app:composite}
To obtain the many-body fractional quantum Hall gaps
and their modification due to the addition of the cavity-mediated interaction,
we make use of the composite fermion trial wavefunctions
for both the ground and excited states.
These wavefunctions are known to be in excellent agreement with the
low-lying exact LLL Coulomb spectrum \cite{jainBook}.
The composite-fermion construction
is based on the idea that the strongly
interacting electron problem at an intense magnetic field can be mapped onto a
weakly interacting problem of composite fermions moving in a reduced effective
magnetic field~\cite{JainCF89}.
Each electron binds to $2p$ fluxes, where $p$ is an integer.
The flux attachment partially screens the external magnetic field, 
so that the effective magnetic field experienced by the composite fermions is
$B^\ast = B - 2p n_{\rm e} \phi_0$.
Equivalently, using $\nu = n_{\rm e}\phi_0/B$ and $\nu^\ast = n_{\rm e}\phi_0/B^\ast$,
the electronic and composite-fermion filling factors are related by
$
    \nu = \nu^\ast / (2p\nu^\ast+1)
$.
For the Laughlin state at $\nu=1/3$, one has $p=1$ and
$\nu^\ast=1$, meaning that the fractional quantum Hall state is described as a completely
filled lowest effective Landau level of composite fermions ($\Lambda$ level),
i.e. an integer quantum Hall state of CFs.
Neutral excitations correspond to promoting a composite fermion to a higher
$\Lambda$ level, forming a composite-fermion exciton, while charge
excitations correspond to creating independent composite-fermion
quasielectrons and quasiholes.

\subsection{Overview of the CF construction on the Haldane sphere}
For our finite-size analysis, we employ the Haldane spherical
geometry~\cite{HaldanePseudoPotentials},
a closed manifold without edges that supports homogeneous states
and preserves rotational invariance,
making it well suited for studying incompressible fractional quantum Hall liquids.
It is particularly convenient as at the appropriate magnetic flux,
the fractional quantum Hall ground state is nondegenerate.

We consider a system of $N$ electrons confined to the surface of a sphere,
subject to a uniform radial magnetic field generated by a magnetic monopole
of strength $Q$ placed at the center.
The monopole strength is constrained by Dirac's quantization condition,
which requires the total magnetic flux through a closed surface,
measured in units of the flux quantum, to be an integer~\cite{DiracQuantization}.
We therefore have
$N_\phi = 2Q \in \mathbb{Z}$, which mandates $R = \ell \sqrt{Q}$ for the sphere's radius.
The filling factor is defined in the thermodynamic limit as
$\nu=N/N_\phi$,
while on the finite sphere, $N$ and $N_{\phi}$ relate as
$N_{\phi} = \nu^{-1}N-S$,
where $S$ is the topological shift,
a quantum number that characterizes the underlying topological order \cite{WenZeeShift}.
For composite fermion states with $2p$ fluxes attached,
in which $\nu^*$ $\Lambda$ levels are filled,
the topological shift is
$S = 2p+ \nu^*$.

The single-particle spectrum on the sphere is obtained by diagonalizing the
kinetic Hamiltonian in the presence of the monopole field.
The eigenstates are the monopole harmonics
$\mathcal{Y}_{Q,n,m}(\Omega)$,
where $\Omega=(\theta,\varphi)$ denotes a point on the sphere~\cite{YangMonopoleHarmonics}.
They are labelled by a Landau-level index $n=0,1,2,\ldots$ and by an orbital angular momentum
quantum number $m=-l_n,\ldots,l_n$, with $l_n = Q+n$.
The corresponding energies are
$
    \epsilon_n
    =
    \hbar^2 / 2m R^2
    \left[
        (Q+n)(Q+n+1)-Q^2
    \right]
$.
The lowest Landau level corresponds to the angular momentum shell
$l_0=Q$, and contains $2Q+1=N_\phi+1$ single-particle states.

Within the composite-fermion construction, the electronic trial wave
functions are obtained by attaching $2p$ fluxes to each electron and
projecting the result to the electronic LLL.
It is convenient to introduce the spinor coordinates
$
    u=\cos(\theta/2)e^{\img\varphi/2}
$
and 
$
    v=\sin(\theta/2)e^{-\img\varphi/2}
$.
The flux attachment factor is
\begin{equation}\label{eqn:vortexFactor}
    J
    =
    \prod_{i<j}
    \left(
        u^{(i)} v^{(j)} - u^{(j)} v^{(i)}
    \right) .
\end{equation}
The general CF trial wave function has the form
\begin{equation}
    \psi^{\rm FQH}_{Q}
    =
    \mathcal{P}_{\rm LLL}
    \left[
        \phi^{\rm CF}_{q}
        J^{2p}
    \right]
    \, ,
\end{equation}
where $\phi^{\rm CF}_{q}$ is a non-interacting wavefunction of
composite fermions moving in an effective monopole field $q$, and
$\mathcal{P}_{\rm LLL}$ denotes projection to the electronic LLL with monopole strength $Q$.
The electron and composite-fermion monopole fluxes
are related by $2q = 2Q - 2p(N-1)$, 
meaning that the flux experienced by a CF is that experienced by an electron, 
subtracted the flux generated by the $2p$ fluxes around each of the $(N-1)$ other particles.

\subsection{Laughlin ground state}
We now specify the CF trial states used throughout this work.
Although the construction can be generalized to all filling factors of the
Jain sequence, $\nu=\nu^*/(2p\nu^*\pm1)$, we focus exclusively
on the Laughlin sequence corresponding to $\nu^*=1$, i.e. $\nu = 1/(2p+1)$.

Let $\mathcal{Y}_{q,n,m}(\Omega)$ denote the monopole harmonic for a composite
fermion at effective monopole strength $q$, $\Lambda$ level index $n$,
and orbital angular momentum $m=-l_n,\ldots,l_n$, with $l_n=q+n$.

The integer quantum Hall ground state of composite fermions
corresponds to a Slater determinant of fully filled $\Lambda$ levels.
For the Laughlin states, this corresponds to the fully filled zeroth $\Lambda$ level
\begin{equation}
    \det_{i,m}
    \left[
        \mathcal{Y}_{q,0,m}(\Omega^{(i)})
    \right],
    \qquad
    m=-q,\ldots,q \; \; ,
\end{equation}
which can be shown to be a Vandermonde determinant,
proportional to the flux attachment factor $J$ defined in Eq.~(\ref{eqn:vortexFactor}).
The lowest $\Lambda$ level therefore contains $2q+1=N$ orbitals.
The electronic ground state is obtained as
$
    \psi^{\nu=1/(2p+1)}_{\rm GS}
    =
    \mathcal{P}_{\rm LLL} \left[
    J \, 
    J^{2p}
    \right]
$.
Since all orbitals lie in the zeroth $\Lambda$ level,
the projection is trivial and one recovers the Laughlin wave function~\cite{Laughlin_wf,HaldanePseudoPotentials}:
\begin{equation}
    \psi^{\nu=1/(2p+1)}_{\rm GS}
    =
    \prod_{i<j}
    \left(
        u^{(i)} v^{(j)} - u^{(j)} v^{(i)}
    \right)^{2p+1} .
\end{equation}

\subsection{CF excitations}
The CF excited states that we employ in this work 
are the independent charged excitations, the quasihole and quasielectron, 
as well as the CF exciton, 
a neutral excitation composed of a bound quasihole-quasielectron pair.

A single quasihole is obtained by increasing the electronic flux by one quantum,
$2Q_{\rm qh}=2Q+1$.
The corresponding CF monopole strength is
$2q_{\rm qh}=2q+1$.
The lowest CF $\Lambda$ level then contains $N+1$ orbitals.
Since there are only $N$ composite fermions, the excitation is represented
as a single hole in an otherwise filled lowest $\Lambda$ level.
A basis of quasihole states is obtained by leaving empty one of the CF orbitals
$m_h=-q_{\rm qh},\ldots,q_{\rm qh}$
\begin{equation}
    \phi^{\rm qh}_{m_h}
    =
    \det_{i,m\neq -m_h}
    \left[
        \mathcal{Y}_{q_{\rm qh},0,m}(\Omega^{(i)})
    \right] \, ,
\end{equation}
forming a degenerate $L=N/2$ multiplet.
The corresponding electronic CF quasihole state is
\begin{equation}
    \psi^{\rm qh}_{m_h}
    =
    \mathcal{P}_{\rm LLL}
    \left[
    \phi^{\rm qh}_{m_h}
    J^{2p}
    \right] \, .
\end{equation}
Again, the projection is trivial, since the composite fermions
occupy only the lowest $\Lambda$ level.

A single quasielectron is obtained by decreasing
the electronic flux by one quantum, $2Q_{\rm qe}=2Q-1$.
The corresponding CF monopole strength is $2q_{\rm qe} = 2q-1$.
The lowest $\Lambda$ level contains $N-1$ orbitals and is filled, while
the remaining composite fermion occupies the first excited $\Lambda$ level.
A basis of quasielectron states is obtained by filling all orbitals of the
lowest CF $\Lambda$ level and placing the remaining composite fermion in 
any orbital of the first excited $\Lambda$ level. Explicitly,
\begin{equation}
    \phi^{\rm qe}_{m_e}
    =
    \det
    \left[
        \left\{
        \mathcal{Y}_{q_{\rm qe},0,m}(\Omega^{(i)})
        \right\} \, , \,
        \mathcal{Y}_{q_{\rm qe},1,m_e}(\Omega^{(i)})
    \right],
\end{equation}
where the first $2q_{\rm qe}+1=N-1$ columns correspond to the filled lowest
$\Lambda$ level, and the last column corresponds to the occupied orbital
$m_e$ in the first excited $\Lambda$ level. The allowed values are
$ m_e=-(q_{\rm qe}+1),\ldots,(q_{\rm qe}+1) $, forming an $L=N/2$ multiplet.
The corresponding electronic quasielectron state is
\begin{equation}
    \psi^{\rm qe}_{m_e}
    =
    \mathcal{P}_{\rm LLL}
    \left[
    \phi^{\rm qe}_{m_e}
    J^{2p}
    \right] \, .
\end{equation}
The LLL projection is implemented using the Jain--Kamilla prescription,
in which the projection operator is brought inside the Slater determinant.
We refer the reader to Ref.~\cite{JainKamilla} for a technical presentation of the method.

Neutral excitations are constructed at the same flux as the ground state.
They correspond to promoting one composite fermion from the highest filled
$\Lambda$ level to the first excited $\Lambda$ level,
thereby creating a CF exciton~\cite{CF_excitons}.
The hole in the lowest $\Lambda$ lives in the $l_h=q$ angular momentum shell,
while the excited composite fermion has angular momentum $l_p=q+1$.
A state with definite total angular momentum $L$
and orbital quantum number $M$
is obtained by coupling the particle and hole angular momenta:
\begin{equation}
    \phi^{\rm exc}_{L M}
    =
    \sum_{m_h,m_p}
    \braket{
        q,-m_h;
        q+1,m_p
        | L,M
    }
    \phi^{\rm exc}_{m_h m_p} \,  .
\end{equation}
Here,
$\braket{q,-m_h;q+1,m_p|L,M}$
is the Clebsch-Gordan coefficient, and
$\phi^{\rm exc}_{m_h m_p}$ is the Slater determinant in which the highest filled
$\Lambda$-level orbital $m_h$ is removed and the first excited
$\Lambda$-level orbital $m_p$ is occupied.
The corresponding electronic composite-fermion exciton is
\begin{equation}
    \psi^{\rm exc}_{L M}
    =
    \mathcal{P}_{\rm LLL}
    \left[
    \phi^{\rm exc}_{L M}
    J^{2p}
    \right] \, .
\end{equation}
The projection is carried out using the Jain--Kamilla method.

Because our Hamiltonian is fully rotationally invariant,
the energies depend only on $L$, not on $M$.
We therefore use the state with $M=0$ throughout.
The spherical angular momentum $L$
plays the role of the neutral excitation momentum.
The planar wave vector recovered in the thermodynamic limit as
$
    k \equiv L/R
$.
We therefore label states by their corresponding wavevector:
\begin{equation}
    \psi^{\rm exc}_{k} \equiv \psi^{\rm exc}_{L=kR,M=0}\, .
\end{equation}
We note that the $L=1$ CF exciton state is annihilated upon projection onto the LLL, 
so the sequence of neutral excitations starts with $L=2$~\cite{CF_excitons, missingStates}.

\section{Computation of many-body energies}\label{app:compuingEnergies}
To compute many-body energies over a continuously varying interaction range,
one would in principle have to evaluate the matrix element
$
    \bra{\psi} \hat{V}_{\rm cav}^{\mathcal{L}} \ket{\psi}
$
by the use of Monte Carlo integration, 
by resampling the many-body states for each pair
potential $V_{\rm cav}(r;\mathcal{L})$.
This is computationally heavy.
Moreover, the Haldane pseudopotentials 
cannot be dealt easily in Monte Carlo 
due to their singular expression in real space.
We use a method introduced in Ref.~\cite{Mross},
which relies on a harmonic decomposition of the interaction
and that has been used to study phase diagrams that depend on the shape of the pair potential.

The angular-momentum-resolved pair density is evaluated once for each state of
interest. The energy for any choice of interaction parameters is then obtained
by contraction with the corresponding pair potential harmonics.
In this section we briefly detail this method.

For any translation and rotation invariant interaction on the sphere, 
the pair interaction between two particles located at 
points $R\bm{\Omega}^{(i)}$ and $R\bm{\Omega}^{(j)}$ depends
only on the squared distance between the points, 
${r_{ij}}^2 = 2 R^2 (1 - x_{ij})$,
with $x_{ij} \equiv \bm{\Omega}^{(i)} \cdot \bm{\Omega}^{(j)} \in [-1,1]$.
This means that we can expand the interaction Hamiltonian in Legendre polynomials
\begin{align}
    \hat{\mathcal H}
    &=
    \frac{1}{2}
    \sum_{i \neq j}
    V(\hat x_{ij})
    =
    \frac{N^2}{2}
    \sum_{L \geqslant 0}
    v_L
    \left(
        \frac{2L+1}{N^2}
        \sum_{i \neq j}
        P_L(\hat x_{ij})
    \right)
    \, ,
\end{align}
with 
\begin{equation}
\label{eqn:harmonicsDef}
V(x)
=
\sum_{L \geqslant 0}
(2L+1)\,
v_L\,
P_L(x)
\end{equation}
defining the pair potential harmonics.
In particular, for the Coulomb interaction, one has~\cite{GMPsphereBalram}:
\begin{equation}
    v^{(\rm C)}_L 
    =
    \frac{\ell}{R}
    \frac{1}{2L+1} \;
    E_{\rm C}
    \, .
\end{equation}

Using the pair harmonics expansion, 
we can cast the expectation value of $\hat{\mathcal H}$ on a generic state $\ket{\psi}$ as:
\begin{equation}\label{eqn:energyGL}
    E_{\psi}
    =
    \frac{N^2}{2}
    \sum_{L \geqslant 0}
    v_L\, G_L(\psi) .
\end{equation}
Here, $G_L(\psi)$ denotes the angular-momentum-resolved pair distribution of the state $\ket{\psi}$:
\begin{align}\label{eqn:Integral}
    \nonumber
    G_L(\psi)
    &\equiv
    \bra{\psi}
    \left(
        \frac{2L+1}{N^2}
        \sum_{i \neq j}
        P_L(\hat x_{ij})
    \right)
    \ket{\psi}
    \\
    &=
    \frac{
        \int \prod_i {\rm d}^2\Omega^{(i)}
        \left(
            \frac{2L+1}{N^2}
            \sum_{i \neq j}
            P_L(x_{ij})
        \right)
        \left|
            \psi(\{\Omega^{(i)}\}_i)
        \right|^2
    }{
        \int \prod_i {\rm d}^2\Omega^{(i)}
        \left|
            \psi(\{\Omega^{(i)}\}_i)
        \right|^2
    }.
\end{align}
We evaluate these integrals using the
Metropolis-Hastings Monte Carlo integration algorithm~\cite{Metropolis, Hastings}
by sampling a weight proportional to $\vert \psi(\{ \Omega^{(i)} \}_i) \vert^2$.
We associate to each evaluated $G_L(\psi)$ a stochastic integration error $\sigma_{G_L}(\psi)$.
As we compute the $\{G_L\}_{L \geqslant 0}$ independently,
the stochastic error of the energy estimator is
\begin{equation}\label{eqn:sigmaEnergy}
    \sigma_E^2
    =
    \sum_{L \geqslant 0}
    \left(\frac{N^2}{2}v_L\right)^2\,\sigma_{G_L}^2 \, .
\end{equation}

Fortunately, for any state in the LLL, one has $G_L(\psi)=0$ for all $L>2Q$.
This follows directly from angular-momentum addition.
Each LLL particle transforms in the monopole-harmonic representation with angular
momentum $Q$. Therefore, a pair of particles belongs to the tensor product
$Q\otimes Q$, which decomposes as
$   
    Q\otimes Q
    =
    \bigoplus_{L=0}^{2Q} L
$.
No two-particle channel with angular momentum larger than $2Q$ exists within
the LLL Hilbert space, and the corresponding pair amplitude must therefore vanish.
This makes the harmonic expansion finite: the pair densities need only be
evaluated for $L=0,\ldots,2Q$.
We hence restrict the sum in Eqs.~(\ref{eqn:energyGL},\,\ref{eqn:sigmaEnergy}) to the said range.

This means that an interaction projected to the LLL,
can be uniquely determined by the set of harmonics $\{ v_L \}_{L=0}^{2Q}$,
which relate to the spherical Haldane pseudopotentials through a linear invertible map 
$v_m = \sum_{L=0}^{2Q} N_{m L} v_L$, with 
\begin{align}
    \label{eqn:vL_to_vm}
    N_{m L}
    &=
    (-1)^{m}(2Q+1)^2
    (2L+1)
    \nonumber\\
    &\quad\times
    \left\{
    \begin{matrix}
        2Q-m & Q & Q \\
        L & Q & Q
    \end{matrix}
    \right\}
    \left(
    \begin{matrix}
        Q & L & Q \\
        -Q & 0 & Q
    \end{matrix}
    \right)^2 \, , 
\end{align}
where $(\dots)$ and $\{\dots\}$ denote the Wigner $3j$ and $6j$
symbols respectively~\cite{Wooten}.

Since we consider spinless electrons throughout this work,
the many-body wave functions are fully antisymmetric under particle exchange.
Consequently, only odd relative-angular-momentum channels contribute,
and the exact interaction energy depends solely on the odd Haldane pseudopotentials.
However, the expression of the Legendre harmonic coefficients ${v_L}$
\emph{does} depend on the even Haldane pseudopotentials ${v_m}$.
Following Ref.~\cite{Mross},
we exploit the freedom of modifying the even pseudopotentials $v_m$,
which leave the exact interaction energy unchanged,
to minimize the stochastic error in Eq.~(\ref{eqn:sigmaEnergy}).
This procedure yields an optimized set of harmonics $\{v_L^{\rm opt}\}$,
which are then used to construct an alternative estimator for the interaction energy with a reduced variance.

\subsection{Harmonics of the cavity-mediated pair-potential}
We defined in Eq.~(\ref{eqnapp:harmonicsCavPlaneCut})
a pair potential in real space on the plane, 
whose Haldane pseudopotentials in the LLL
correspond to the ones for the cavity-mediated interaction.
On the sphere, the Haldane pseudopotentials are different
and they form a finite set.
In this section we derive the pair potential harmonics
for both the infinite range potential from Eq.~(\ref{eqnapp:harmonicsCavPlane}) 
and the finite range one, given in Eq.~(\ref{eqnapp:harmonicsCavPlaneCut}).

We will assume that, on the sphere, particles at a chord distance $r$
interact through the same pair potential $V_{\rm cav}(r;\mathcal{L})$ as on the plane.
In order to get the pseudopotentials, we need to first identify the harmonics $v^{\rm (cav)}_L$.
We start by expressing the cavity-mediated interaction $V_{\rm cav}(r)$
in terms of the angular separation on the sphere $x$. 
By using $r^2 = 2R^2 (1-x)$
and the definitions of the first three Legendre polynomials, 
we can show that:
\begin{align}
    \nonumber
    V_{\rm cav}(r)
    &=
    -\xi
    \Bigg[
        \left(
            \frac{R^4}{3\ell^4}
            -
            \frac{2R^2}{\ell^2}
            +
            2
        \right)
        P_0(x)
        \\&\quad+
        3
        \left( 
            \frac{2R^2}{3\ell^2} - \frac{R^4}{6\ell^4}
        \right)
        P_1(x)
        +
        5
        \left(
            \frac{R^4}{30\ell^4}
        \right)
        P_2(x)
    \Bigg]
    \, .
\end{align}
With this, we read out the harmonics
from the expression:
\begin{equation}
    V_{\rm cav}(r) = \sum_{L \geqslant 0} v^{\rm (cav)}_L (2L+1) P_L(x) \, .
\end{equation}

The harmonics for the finite range $V_{\rm cav}(r; \mathcal{L})$
can be obtained from $ v^{\rm (cav)}_L$ by convoluting them to the harmonics of the Gaussian,
which can be shown to be:
\begin{equation}
    e^{-r^2/\mathcal{L}^2}
    =
    \sum_{L \geqslant 0}
    (2L+1)
    \left[
        i_L(\alpha)
        e^{-\alpha}
    \right]
    P_L(x) \, .
\end{equation}
Here, $\alpha \equiv 2R^2 / \mathcal{L}^2$
and $i_L(\alpha)$ is the modified spherical Bessel function of the first kind \cite{dlmf}.
The harmonic component of an interaction is obtained 
by integrating $V(x)P_L(x)/2$ over the interval $[-1,1]$.
To determine the harmonics of $V_{\rm cav}(r;\mathcal{L}) = V_{\rm cav}(r)e^{-r^2/\mathcal{L}^2}$,
we first expand both factors in Legendre polynomials
and perform the integral analytically.
The integral involves three Legendre polynomials
and can be evaluated using the standard Gaunt relation.
The result is given by:
\begin{align}
    \nonumber
    v^{(\rm cav)}_L(\mathcal{L})
    &=
    \sum_{L'=0}^{2Q}
    \sum_{L''=0}^{2Q}
    (2 L'+1)
    (2 L'' + 1) 
    \\ &\quad\quad\times
    v_{L'}^{(\rm cav)} 
    \left[
        i_{L''}(\alpha)
        e^{-\alpha}
    \right]
    \begin{pmatrix}
        L & L' & L'' \\
        0&0&0
    \end{pmatrix}^2 \, .
\end{align}

\section{\label{app:ED_benchmarks}
    Benchmarks with exact diagonalization results
}

\begin{figure*}[t!]
    \centering
    \resizebox{0.85\textwidth}{!}{%
        \includegraphics[width=\columnwidth]{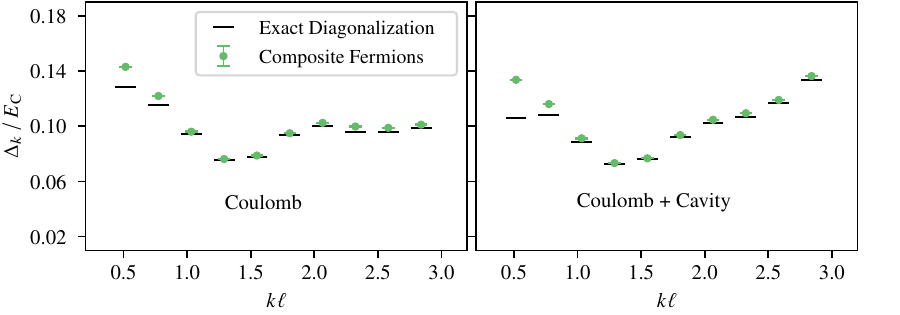}
    }
    \caption{
        Neutral excitation gaps at filling factor $\nu=1/3$
        obtained from exact diagonalization
        and from the composite-fermion exciton ansatz.
        Bare Coulomb interaction (left panel)
        and cavity-modified spectrum (right panel).
        The results are shown for $N=11$ electrons,
        $\mathcal L/\sqrt{\mathcal{S}}=\infty$,
        and $\xi N^2=0.23E_{\rm C}$.
        Black lines represent the ED spectra, 
        whereas green dots, the CF results.
    }
    \label{fig:ED_v_CF_neutral}
\end{figure*}

In this appendix we present some exact diagonalization (ED) results that 
serve as a benchmark for the composite fermion theory at small system sizes.
We begin by detailing the ED procedure and then report comparisons between ED results
and those obtained with CFs.

Any two-body interaction within the LLL on the sphere, 
written in second quantization reads:
\begin{equation}
    \hat{\mathcal{H}}
    =
    \frac{1}{2}
    \sum_{m_1 m_2 m'_1 m'_2}
    \bra{m_1;m_2}
        \hat{\mathcal{H}}
    \ket{m'_1;m'_2}
    \hat{c}^{\dagger}_{m_1}
    \hat{c}^{\dagger}_{m_2}
    \hat{c}_{m'_2}
    \hat{c}_{m'_1} \, ,
\end{equation}
where $\hat{c}^{\dagger}_{m}$ is the creation operator of 
the orbital $\mathcal{Y}_{Q,0,m}$, with $m \in \{-Q,\dots,+Q\}$,
and 
$
    \bra{m_1;m_2}
        \hat{\mathcal{H}}
    \ket{m'_1;m'_2}
$
are the matrix elements in the two-body product basis.
For any translation and rotation invariant pair potential,
these matrix elements can be entirely characterized
by the Haldane pseudopotentials $\{v_m\}_{m=0}^{2Q}\,$,
\begin{align}
    \nonumber
    \bra{m_1;m_2}
        \hat{\mathcal{H}}
    \ket{m'_1;m'_2}
    &=
    \sum_{L=0}^{2Q}
    \sum_{M=-L}^{+L}
    v_{2Q-L} 
    \langle Q m_1,Q m_2 \vert LM \rangle
    \\
    &\times
    \langle LM \vert Q m'_1,Q m'_2\rangle \, ,
\end{align}
where the matrix elements involved above are the Clebsch-Gordan coefficients.
\begin{equation}
    v^{\rm (C)}_{m}
    =
    \frac{2 \ell}{R}
    \frac{
        \binom{2m}{m}
        \binom{8Q-2m+2}{4Q-m+1}
    }{
        \binom{4Q+2}{2Q+1}^2
    }
    E_{\rm C}
    \, ,
\end{equation}
are the Coulomb Haldane pseudopotentials~\cite{FanoOrtolaniColombo}.
We evaluate charged excitation energies by fixing the spherical radius.
We therefore distinguish the radius parameter $(R/\ell)^2$
from the monopole strength $Q$.
In the ground-state sector $(R/\ell)^2=Q$,
whereas in the charged sectors $Q=(R/\ell)^2\pm 1/2$.

In our ED we employ the $\mathcal{L}/\sqrt{\mathcal{S}} \to +\infty$ potential, $V_{\rm cav}(r)$.
We obtain the following pseudopotentials
\begin{align}
    \nonumber
    v_m^{\rm (cav)}
    &=
    \xi \Big[
    -2
    +
    4 \frac{R^2}{\ell^2}
    \frac{(m+1)(4Q-m+2)}{(2Q+2)^2}
    \\
    &
    -
    \frac{R^4}{\ell^4}
    \frac{(m+1)(m+2)(4Q-m+2)(4Q-m+3)}
    {(2Q+2)^2(2Q+3)^2}
    \Big] \, .
\end{align}
We note in passing that the result can be obtained by computing
$\sum_L N_{m L} v_L^{(\rm cav)}$; see Eq.~(\ref{eqn:vL_to_vm}).

To perform exact diagonalization,
we first construct the many-body Fock basis
and then build the Hamiltonian matrix in this basis.
Since the Hamiltonian is rotationally invariant,
$[\hat{\mathcal{H}},\hat{L}_z]=[\hat{\mathcal{H}},\hat{L}^2]=0$,
the spectrum decomposes into angular-momentum multiplets
labelled by the total angular momentum $L$.
In practice, we exploit only the conservation of $L_z$ at the level of the basis construction
and diagonalize the Hamiltonian within a fixed $L_z$ sector.
For the neutral spectrum, we work in the $L_z=0$ sector, which contains one representative of each multiplet.
After diagonalization, the value of $L$ associated with each eigenstate $\ket{\psi}$
is obtained by evaluating the expectation value of $\bra{\psi}\hat{L}^2\ket{\psi} = L(L+1)$.
The ground state is non-degenerate and has $L=0$.

For the quasihole and quasielectron sectors relevant to the charge gap,
assuming the cavity does not alter the structure of the FQH phase,
the lowest-energy states form a multiplet with $L=N/2$.
We therefore carry out the diagonalization in the $L_z=N/2$ sector,
where this multiplet has a single highest-weight representative,
and extract the corresponding non-degenerate ground-state.

\subsection{Neutral Spectrum}

In Fig.~\ref{fig:ED_v_CF_neutral},
we show a comparison between the neutral gaps obtained through ED and the CF exciton ansatz
at filling $\nu=1/3$.
For ED, we diagonalize the full Hamiltonian:
\begin{equation}
    \hat{\mathcal{H}} = \hat{V}^{\rm LLL}_{\rm C} + \hat{V}_{\rm cav} \, .
\end{equation}
The spectrum splits into angular momentum multiplets
and we report the lowest lying energy branch for $L=2,\dots,N$,
which we translate to the planar wavevector $k \equiv L/R$.

As is evident from Fig.~\ref{fig:ED_v_CF_neutral}, the exact results are very well captured by the CF ansatz.
In particular, the CF spectrum closely follows the exact Coulomb spectrum,
with deviations of only a few percent.

We emphasize, however, that the CF ansatz should be viewed as a variational
approximation to the exact Coulomb eigenstates, rather than as the exact
eigenbasis of a microscopic Hamiltonian. Consequently, the energy scale
$A_{\nu}(\mathcal{L}/\sqrt{\mathcal{S}}) \xi N^2$ that controls the gap shift,
should remain small compared to the Coulomb gap,
which is of order $0.1\,E_{\rm C}$. In this regime, the CF ansatz provides
a reliable first-order estimate of the correction to the excitation gap.

\subsection{Charge gap shift}

\begin{figure}[t!]
    \centering
    \includegraphics[width=\columnwidth]{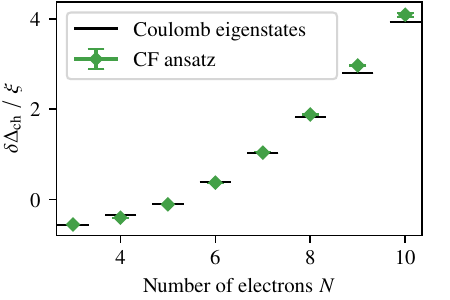}
    \caption{
        Cavity-induced correction to the charge gap of the $\nu=1/3$ state,
        as a function of the number of electrons.
        A comparison between results obtained with the composite fermion wavefunctions
        and the Coulomb eigenstates from exact diagonalization.
        To be compared to the large scale simulations reported in Fig.~\ref{fig:chargeGapShiftsInfCutOff_vs_N_manyLs}.
    }
    \label{fig:ED_chGapShifts}
\end{figure}

In Fig.~\ref{fig:ED_chGapShifts},
we show how the charge gap shift induced by the cavity.
We diagonalize the Coulomb Hamiltonian in the ground state,
quasi-hole and quasi-electron sectors
and we obtain the corresponding ground states 
$
    \ket{\psi^{(\rm C)}_{\rm gs}},
    \ket{\psi^{(\rm C)}_{\rm qh}},
    \ket{\psi^{(\rm C)}_{\rm qe}}
$.
We then compute the expectation value of the cavity-mediated interaction $\hat{V}_{\rm cav}$
onto these values and plot
\begin{equation}
    \bra{\psi_{\rm qh}^{\rm (C)}} \hat{V}_{\rm cav} \ket{\psi_{\rm qh}^{\rm (C)}} 
    +
    \bra{\psi_{\rm qe}^{\rm (C)}} \hat{V}_{\rm cav} \ket{\psi_{\rm qe}^{\rm (C)}} 
    -
    2
    \bra{\psi_{\rm gs}^{\rm (C)}} \hat{V}_{\rm cav} \ket{\psi_{\rm gs}^{\rm (C)}} 
\end{equation}
as a function of the number of particles
and compare to the result obtained using the CF ansatz states for the 
quasi-hole, quasi-particle and the ground state, see Eq.~(\ref{eqn:chGap_shift}).
The relative agreement is within $\sim 0.1-1 \%$,
for system sizes up to $N=10$ electrons.
This agreement demonstrates that the CF wave functions
provide highly accurate approximations to the exact Coulomb eigenstates,
and serves as a stringent validation of our numerical implementation.

For very small number of electrons, $N<6$, we find a decrease of the charge gap
rather than an increase.
We attribute this behavior to finite-size effects on the sphere:
at these system sizes, the spherical pseudopotentials associated with
the cavity-mediated interaction switch relative signs and therefore no longer
provide a reliable representation of their planar counterparts, even at
short distances.

\subsection{Thermodynamic Extrapolations of the charge gap}

\begin{figure}[t!]
    \centering
    \includegraphics[width=\columnwidth]{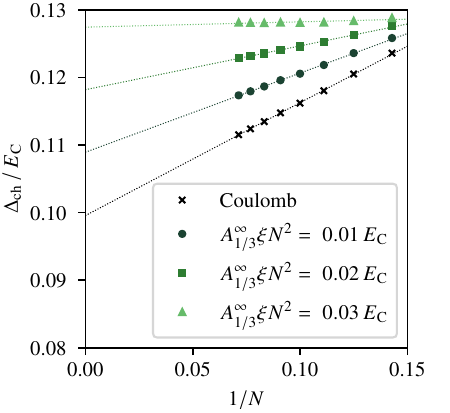}
    \caption{
        Thermodynamic extrapolation of the charge gap
        of the Laughlin $\nu=1/3$ state obtained through ED
        for an interaction range $\mathcal{L}/\sqrt{\mathcal{S}} \to +\infty$, 
        with gaps being shown as a function of $1/N$ and fitted to a straight line,
        for $N=7,\dots,14$.
        To extract this thermodynamic behavior of the cavity-modified gaps,
        $\xi$ has been adjusted to keep $\xi N^2$ fixed, as in Fig.~\ref{fig:cavity_tl_CF}.
        We report the coupling strength as $A^{\infty}_{1/3} \xi N^2$, where
        $
            A^{\infty}_{1/3}
            \equiv
            A_{1/3}(\mathcal{L}/\sqrt{\mathcal{S}} \to +\infty)
            =
            0.087
        $.
    }
    \label{fig:cavity_tl_ED}
\end{figure}

In the main text, the thermodynamic extrapolation of the cavity-modified charge gap
was obtained using composite-fermion wave functions,
which allow us to reach larger system sizes.
Here, we repeat the same analysis using ED
for the smaller systems where it is accessible.

As in the composite-fermion calculation,
we focus on the limit $\mathcal{L}/\sqrt{\mathcal{S}}\to+\infty$,
where the cavity-mediated interaction is effectively all-to-all.
To obtain a well-defined thermodynamic extrapolation,
for each value of $N$ we tune $\xi$ such that the scaled coupling
$\xi N^2$ remains fixed.

The resulting charge gaps are shown in Fig.~\ref{fig:cavity_tl_ED} as a function of $1/N$.
For each value of $A^{\infty}_{1/3}\xi N^2$,
we fit the finite-size data to a linear form
in $1/N$ and extract the intercept at $1/N=0$.
The extrapolated gaps increase by an amount $\approx A^{\infty}_{1/3}\xi N^2$,
in agreement with the composite-fermion extrapolation shown in
Fig.~\ref{fig:cavity_tl_CF}.
This provides an independent exact-diagonalization benchmark
for the scaling
procedure used in the main text.

\section{\label{app:nu_15}
    Finite size scaling for the $\nu=1/5$ state charge gap
}
In the main text, we presented a finite-size scaling analysis
of the cavity-induced charge-gap shift for the Laughlin
$\nu=1/3$ state. In Fig.~\ref{fig:chargeGapShiftsInfCutOff_vs_N_manyLs},
we showed this shift as a function of the number of electrons
for several values of the rescaled cutoff length
$\mathcal{L}/\sqrt{\mathcal{S}}$.
The data were found to be well described by a quadratic dependence on
system size, which allowed us to extract the coefficient
$A_{1/3}(\mathcal{L}/\sqrt{\mathcal{S}})$.

\begin{figure}[t!]
    \centering
    \includegraphics[width=\columnwidth]{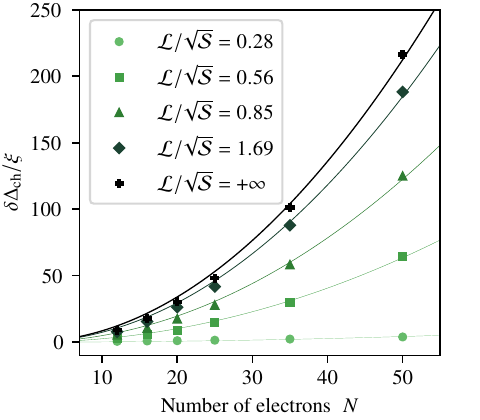}
        \caption{
            Cavity-mediated variation of charge-gap $\delta\Delta_{\mathrm{ch}}/\xi$
            as a function of the number of electrons $N$
            for the $\nu=1/5$ fractional quantum Hall state
            and for different values of $\mathcal{L}/\sqrt{\mathcal{S}}$.
            Symbols denote numerical results, while solid lines are fits to
            $
                \delta\Delta_{\mathrm{ch}} 
                =
                A_{1/5}(\mathcal{L}/\sqrt{\mathcal{S}})\,\xi N^2
            $
            where $A_{1/5}(\mathcal{L}/\sqrt{\mathcal{S}})$ depends on the interaction range.
            The fit is performed using all the data points on the plot.
            The increase of $A_{1/5}$ with $\mathcal{L}/\sqrt{\mathcal{S}}$
            reflects the increasingly long-range character of the cavity-mediated interaction.
            The characteristic energy scale $\xi$,
            defined in Eq.~(\ref{eqn:xi}),
            is proportional to the fourth power of the cavity-mode spatial gradient.
    }
    \label{fig:chargeGapShiftsInfCutOff_vs_N_manyLs_15}
\end{figure}

As stated in the main text, an analogous analysis was performed for the
Laughlin $\nu=1/5$ state and was used in the construction of
Fig.~\ref{fig:delta13_delta15}. For completeness, we present in Fig.~\ref{fig:chargeGapShiftsInfCutOff_vs_N_manyLs_15},
the corresponding finite-size scaling data for $\nu=1/5$,
analogous to Fig.~\ref{fig:chargeGapShiftsInfCutOff_vs_N_manyLs} in the main text.

\section{\label{app:PseudoPotentials}
    Cavity-induced Haldane pseudopotentials on the plane
}
To obtain the Haldane pseudopotentials
of the cavity pair potential, given in Eq.~(\ref{eqn:cavity_plane_pps}),
we compute the interaction energy of a pair of particles
at fixed relative angular momentum $m$.
We can express the corresponding state, $\ket{M=0,m}$,
in terms of the two-body product states
by applying the binomial expansion as follows:
\begin{align}
    \ket{M=0,m}
    &=
    \sum_{p=0}^m
    \frac{(-1)^p}{\sqrt{2^m}} \sqrt{ \binom{m}{p} } 
    \ket{m-p;p} 
    \, .
\end{align}
We then compute the following matrix element
that gives the Haldane pseudopotential:
\begin{align}
    \nonumber
    v_m
    &=
    -\xi'    
    \bra{M=0,m}
    \hat{X}^{(1)}
    \hat{X}^{(2)}
    \ket{M=0,m}
    \\
    \nonumber
    &=
    -\xi'    
    \sum_{p=0}^m
    \sum_{q=0}^m
    \frac{(-1)^{p+q}}{2^m}
    \sqrt{ 
        \binom{m}{p}
        \binom{m}{q}
    }
    \\
    &\times
    \bra{m-q} \hat{X} \ket{m-p}
    \bra{q} \hat{X} \ket{p} \, .
\end{align}
We have used the fact that 
$\hat{X}^{(1)} \hat{X}^{(2)}$ factorizes into single particle operators.
We introduced here $\xi' \equiv 2 \xi / 3$.
The expression for $\bra{r} \hat{X} \ket{s}$ is readily obtained
by employing the ladder structure of the operator 
$\hat{X} = 2 b^{\dagger}b + b^2 + {b^{\dagger}}^2$.
We obtain:
\begin{align}
    \nonumber
    \bra{r} \hat{X} \ket{s}
    &=
    2s \,\delta_{r,s}
    +
    \sqrt{s(s-1)} \,\delta_{r,s-2}
    \\&+
    \sqrt{(s+1)(s+2)} \,\delta_{r,s+2} \, .
\end{align}
The matrix elements of $\hat{X}$ only connect states with angular momentum
difference $0$ or $\pm 2$.
Accordingly, only terms with $r=s$, $r=s+2$, or $r=s-2$ contribute to the sum.
We therefore decompose
\begin{equation}
    v_m = v_m^{(0)} + v_m^{(+2)} + v_m^{(-2)} \, .
\end{equation}
The diagonal contribution, corresponding to $q=p$, is
\begin{align}
    \nonumber
    v_m^{(0)}
    &=
    -\xi'
    \sum_{p=0}^m
    \frac{1}{2^m}
    \binom{m}{p}
    \,
    2(m-p)\, 2p
    \\
    &=
    -\frac{4\xi'}{2^m}
    \sum_{p=0}^m
    \binom{m}{p}
    p(m-p) \, .
\end{align}
Using the binomial identities
\begin{align}
    \sum_{p=0}^m \binom{m}{p} p &= m 2^{m-1} \, ,\\
    \sum_{p=0}^m \binom{m}{p} p^2 &= m(m+1) 2^{m-2} \, ,
\end{align}
we obtain
\begin{equation}
    v_m^{(0)} = -\xi' \, m(m-1) \, .
\end{equation}
The off-diagonal contribution with $q=p+2$ reads
\begin{align}
    \nonumber
    v_m^{(+2)}
    &=
    -\xi'
    \sum_{p=0}^{m-2}
    \frac{1}{2^m}
    \sqrt{
        \binom{m}{p}
        \binom{m}{p+2}
    }
    \\
    &\times
    \sqrt{(m-p)(m-p-1)}
    \sqrt{(p+1)(p+2)} \, .
\end{align}
Using
\begin{equation}
    \binom{m}{p+2}
    =
    \binom{m}{p}
    \frac{(m-p)(m-p-1)}{(p+1)(p+2)} \, ,
\end{equation}
this simplifies to
\begin{align}
    \nonumber
    v_m^{(+2)}
    &=
    -\frac{\xi'}{2^m}
    \sum_{p=0}^{m-2}
    \binom{m}{p}
    (m-p)(m-p-1)
    \\
    &=
    -\frac{\xi'}{2^m}
    \sum_{p=0}^{m}
    \binom{m}{p}
    (m-p)(m-p-1) \, .
\end{align}
By symmetry of the binomial distribution,
\begin{align}
    \nonumber
    &
    \sum_{p=0}^{m}
    \binom{m}{p}
    (m-p)(m-p-1)
    \\&=
    \sum_{p=0}^{m}
    \binom{m}{p}
    p(p-1)
    =
    m(m-1) 2^{m-2} \, ,
\end{align}
and therefore
\begin{equation}
    v_m^{(+2)} = -\frac{\xi'}{4} m(m-1) \, .
\end{equation}
By the same reasoning,
\begin{equation}
    v_m^{(-2)} = -\frac{\xi'}{4} m(m-1) \, .
\end{equation}
Summing all contributions, we finally obtain
\begin{equation}
    v_m
    =
    -\frac{3}{2}\,\xi' \, m(m-1)
    =
    -\xi (m^2 - m)
    \, .
\end{equation}

\bibliography{./Bib_.bib}

\end{document}